\documentclass[useAMS,usenatbib, usegraphicx]{mn2e}

\usepackage{graphicx}
\usepackage{hyperref}
\usepackage{amsmath,amssymb}
\usepackage{amsfonts}
\usepackage{xspace} 
\usepackage[usenames]{color}
\usepackage{dcolumn} 
\usepackage{bm} 
\usepackage{epsfig}
\usepackage{aas_macros}

\usepackage{ulem}
\definecolor {darkgreen}{rgb}{0.2,0.7,0.2}

\newcommand{\be}{\begin{equation}}
\newcommand{\ba}{\begin{eqnarray}}
\newcommand{\ee}{\end{equation}}
\newcommand{\ea}{\end{eqnarray}}

\newcommand{\SMBH}{\bullet}
\newcommand{\SCO}{\rm s}

\newcommand{\GW}{{\mbox{\tiny GW}}}

\newcommand{\ISCO}{{\mbox{\tiny ISCO}}}

\newcommand{\Msun}{\,{\rm M_\odot}}

\newcommand{\cm}{\,{\rm cm}}

\newcommand{\D}{\mathrm{d}}
\newcommand{\C}{c}

\newcommand{\R}{{\bar{r}}}

\newcommand{\G}{G}
\newcommand{\Beta}{B}
\newcommand{\W}{\mathcal{W}}

\title[Gas pile-up, gap overflow, and Type 1.5 migration: general theory]{Gas pile-up, gap overflow, and Type 1.5 migration in circumbinary disks: general theory}

\author[Kocsis, Haiman, \& Loeb]{Bence Kocsis$^{1,3}$\thanks{E-mail: bkocsis@cfa.harvard.edu},
Zolt\'an Haiman$^{2}$\thanks{E-mail: zoltan@astro.columbia.edu}
and Abraham Loeb$^{1}$\thanks{E-mail: aloeb@cfa.harvard.edu}\\
$^{1}$Harvard-Smithsonian Center for Astrophysics, 60 Garden St., Cambridge, MA 02138, USA\\
$^{2}$Department of Astronomy, Columbia University, 550 West 120th Street, New York, NY 10027\\
$^{3}$Einstein Fellow}

\begin{document}
\maketitle

\begin{abstract}
Many astrophysical binaries, from planets to black holes, exert strong
torques on their circumbinary accretion disks, and are expected to
significantly modify the disk structure. Despite the several decade
long history of the subject, the joint evolution of the binary + disk
system has not been modeled with self-consistent assumptions for
arbitrary mass ratios and accretion rates.  Here we solve the coupled
binary-disk evolution equations analytically in the strongly perturbed
limit, treating the azimuthally-averaged angular momentum exchange
between the disk and the binary and the modifications to the density,
scale-height, and viscosity self-consistently, including viscous and
tidal heating, diffusion limited cooling, radiation pressure, and the
orbital decay of the binary.  We find a solution with a { central cavity}
and a migration rate similar to those previously obtained for Type-II
migration, applicable for large masses and binary separations, and
near-equal mass ratios.  However, we identify a distinct new regime,
applicable at smaller separations and masses, and mass ratio in the
range $ 10^{-3}\lesssim q \lesssim 0.1$.  For these systems, gas piles
up outside the binary's orbit, but rather than creating a cavity, it
continuously overflows as in a porous dam.  The disk profile is
intermediate between a weakly perturbed disk (producing Type-I
migration) and a disk with a gap (with Type-II migration). However,
the migration rate of the secondary is typically slower than both
Type-I and Type-II rates.  We term this new regime ``Type-1.5''
migration.
\end{abstract}


\section{Introduction}

Understanding the co-evolution of binaries and accretion disks is
fundamental in several fields of astrophysics, including planet
formation and migration
\citep{1980ApJ...241..425G,1997Icar..126..261W}, patterns in planetary
rings \citep{1982ARA&A..20..249G}, stellar binaries, compact object,
and binaries involving supermassive black holes (SMBHs).

Despite the long history of the subject, there are no self-consistent
analytical models for the co-evolution of binaries and accretion
disks, incorporating the fundamental physical effects over the long
timescales on which the binary separation evolves.  The standard
$\alpha$--model of radiatively efficient turbulent thin accretion
disks \citep{1973A&A....24..337S} relates the effective kinematic
viscosity of the disk to the pressure $\nu \propto \alpha p$.  The viscous
evolution of the disk, however, is often modeled without considering
the pressure dependence of the viscosity
\citep{1974MNRAS.168..603L}. Similarly, models of the gravitational
interaction between the disk, which describe the launching of spiral
density waves in the disk that remove angular momentum from the
binary, also do not account for the tidal heating of the disk and the
corresponding feedback on the torque cutoff phenomenon
\citep{1980ApJ...241..425G}.

The evolution of the circumbinary disk is sensitive to the above
mentioned assumptions, especially when the mass of the secondary is
large, and can strongly perturb the disk.  For a massive secondary,
the tidal torque clears a gap in the disk, and the viscous radial
inflow of the gas pushes the object inward on the viscous timescale
(Type-II migration).  If the secondary mass, $m_{\SCO}$ is larger than
the local disk mass, $m_{\rm d}=4\pi r^2 \Sigma$, where $\Sigma$ is
the surface density, then the migration slows down, as the spiral
density waves cannot remove angular momentum away from the binary at a
rate on which the gas flows in.  This leads to the pile-up of gas
outside the secondary's orbit, in which the gas density increases by { up to} a
factor $B^{-3/8}$, where $B=m_{\rm d 0}/m_{\SCO}<1$ and $m_{\rm d0}$
is the unperturbed local disk mass \citep{1995MNRAS.277..758S}. Once
this steady-state level is reached, the viscous gas inflow velocity
matches the inward migration of the object (secondary dominated
Type-II migration).

In this paper, we focus on such systems, with $m_{\SCO}>m_{\rm d 0}$,
and point out that the \citet{1995MNRAS.277..758S} steady-state level
of gas pile-up cannot be reached for sufficiently large secondary
masses $m_{\SCO} \gg m_{\rm d 0}$, where $B\ll 1$.  The enhanced
viscous dissipation rate ($D_{\nu}\propto B^{-5/8}$) can increase the
disk temperature such that it becomes radiation pressure
dominated. The enhanced pressure makes the disk puff up ($H\propto
B^{-5/8}$), and reduces the relative gap size. Once the gas approaches
within a distance less than the scale-height from the secondary, the
torque that the disk exerts on the binary has a cutoff
\citep{1980ApJ...241..425G,1993ApJ...419..155A,1993ApJ...419..166A,2001ApJ...552..793G}
which limits the migration rate of the secondary.  Once the gas enters
the Hill radius, it can furthermore flow across the secondary's orbit
along horse shoe orbits or accrete onto the secondary.
We derive an analytical { quasi-}steady-state model for the co-evolution of the
disk and the orbital migration of the secondary, in which we combine a
\citet{1973A&A....24..337S} disk with the theory of the binary-disk
interaction by \citet{1980ApJ...241..425G} self-consistently.  In
particular, we adopt the viscosity prescription of standard thin
accretion disks proportional to pressure,\footnote{ Here $\nu\propto
p_{\rm gas}$ and $\nu\propto (p_{\rm gas}+p_{\rm rad})$ are
respectively known as $\alpha$ and $\beta$-models.  We formulate our
problem for a general $\alpha$ or $\beta$-disk in Sec.~2, but then
derive the analytical results for the special case of a $\beta$ disk
viscosity.}  calculate the sound speed and vertical balance including
both gas and radiation pressure ($p_{\rm gas}$ and $p_{\rm rad}$),
adopt the simple analytical approximation to the angular momentum
exchange between the binary and the disk of
\citet{2002ApJ...567L...9A}, consider the standard viscous and tidal
heating of the disk \citep{2009MNRAS.398.1392L}, and self-consistently
account for the feedback on the pressure, viscosity, scale-height, and
the torque cutoff near the secondary's orbit.
{We generalize
the steady-state model of \citet{1984Icar...60...29H}, \citet{1989ApJ...347..490W}, and \citet{2010PhRvD..82l3011L}
by self-consistently including variations in the viscosity and pressure caused by the pile-up.}
We derive azimuthally
averaged { steady-state} analytical disk models which recover the
\citet{2004ApJ...608..108G} solution for arbitrary $\beta=p_{\rm
gas}/(p_{\rm gas}+p_{\rm rad})$ in the limit that the secondary mass
$m_{\SCO}$ approaches zero, but the disk structure is significantly
modified by the secondary over multiple accretion timescales for
larger $m_{\SCO}$.

{ The disk structure in this overflowing state with a pile-up is intermediate between
the weakly perturbed case without a secondary and the case with a gap.
Not surprisingly, the migration rate in such an intermediate state,
which we label Type-1.5, is significantly different from the corresponding limiting cases
of Type-I migration and the secondary-dominated Type-II migration.
The transition between Type-I and Type-II migration as a function of the secondary mass
was previously typically investigated by considering only the change
in the surface density due to gap formation, but without investigating
the feedback from the changes in viscosity and pressure
\citep{1984Icar...60...29H,1989ApJ...347..490W,1996ApJS..105..181K,1997Icar..126..261W,2003MNRAS.341..213B,2007MNRAS.377.1324C}.
However, simulations show that migration is sensitive to temperature variations and
radiation pressure
\citep{2003ApJ...599..548D,2006A&A...459L..17P,2008A&A...478..245P,2008A&A...487L...9K}.
We derive the Type-1.5 migration rate for the self-consistent radial
profile including these effects when the pile-up is significant
in an overflowing steady-state disk. 
As the secondary migrates inwards across the increasingly hotter inner regions of the disk, 
the gap opening conditions and the migration rate change even if one 
neglects the feedback on viscosity and temperature due to gas pile-up 
\citep{2009CQGra..26i4032H,2011PhRvD..84b4032K}, but here we show that 
the changes are significant over a much wider range of masses and radii in the self-consistent model.
We discuss migration and gap opening
for SMBH binaries in more detail in \citet{paper2}, hereafter Paper~II.
}

The remainder of the paper is organized as follows.
In \S~\ref{s:interaction}, we lay out the basic equations governing
the hydrodynamical and thermal evolution of the disk, as well as the
migration of the secondary. We solve the equations numerically in \S~\ref{s:numerical}.
We then derive an analytical solution in
\S~\ref{s:app:analytical}. We summarize the results, and discuss how
they depend on the most important physical parameters, in
\S~\ref{s:analytical}.  We offer our conclusions in
\S~\ref{s:conclusions}.  A more detailed discussion and the implications
for SMBH binary systems is presented in { Paper~II}.

\begin{figure}
\centering
\mbox{\includegraphics[width=8.5cm]{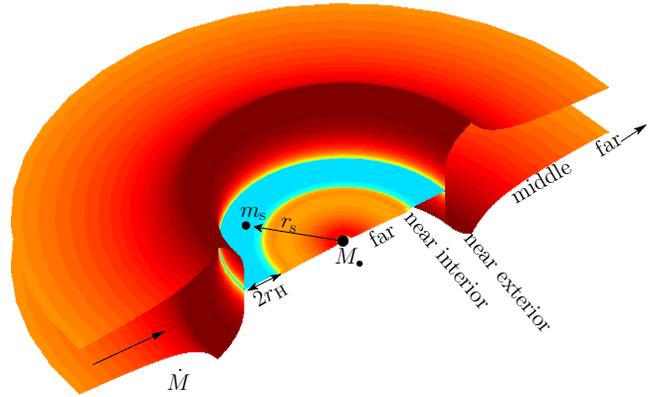}}
\caption{\label{f:disk}
Gas pile-up and overflow in a circumbinary accretion disk with
component masses $M_{\SMBH}$ and $m_{\SCO}$, binary separation
$r_{\SCO}$, and accretion rate $\dot{M}$. We distinguish five distinct
radial zones: an inner and an outer far zone where the effects of the
secondary are negligible, an interior and an exterior near zone where
the tidal effects are significant, and an extended middle zone with a
significant gas pile-up (see \S~\ref{s:app:analytical}).  }
\end{figure}

We use geometrical units $\G=\C=1$, and suppress factors of $\G/\C^2$
and $\G/\C^3$ to convert between mass, length, and time units. Our
basic notation for the disk and secondary parameters are depicted in
Figure~\ref{f:disk}.

\section{Thermo-hydrodynamical interaction between a disk and a secondary}\label{s:interaction}

We examine the evolution of the secondary and an azimuthally and
vertically averaged Shakura-Sunyaev disk (i.e. axisymmetric one-zone
disk) in local thermal equilibrium.  Here we review the basic
equations.  First, we write down the continuity and angular momentum
transport equations including the viscous torque and the gravitational
tidal torque of the secondary.  The back-reaction of the tidal torque
changes the angular momentum of the secondary.  The viscous and tidal
torques depend on the disk surface density, viscosity, and pressure
gradient (or scale-height). We derive the vertically averaged disk
structure assuming that: (i) the local viscous plus tidal heating
equals the radiative cooling with photon diffusion limited vertical
radiative flux (i.e. negligible radial heat transport); (ii) the
viscosity is proportional to either gas+radiation pressure
($\alpha$-disk) or just the gas pressure ($\beta$-disk), and (iii) gas
plus radiation pressure supports the disk against the vertical
gravity.  This yields a closed set of nonlinear partial differential
equations for the disk and the location of the secondary in 1+1
dimensions. We present solutions in subsequent sections below.

\subsection{Angular momentum transport}

We denote the masses of the primary and the secondary objects by
$M_{\SMBH}$ and $m_{\SCO}$, the surface density of the disk by
$\Sigma$ (assuming axisymmetry), and the radial bulk velocity of the
disk by $v_r$, which is negative if gas accretes toward $r=0$.  The
continuity and angular momentum equations for the disk are\footnote{In
our notation, $\partial_r \Sigma \equiv \partial \Sigma / \partial r$
and $\partial_t \Sigma \equiv \partial \Sigma / \partial t \equiv
\dot{\Sigma}$.
{ $T$ refers to torque, $T_{\nu}$ and $T_{\rm d}$ are viscous and tidal torques
as in \citet{2010MNRAS.407.2007C,2010PhRvD..82l3011L}. The angular momentum flux is
$F_{\rm J}\equiv T$.
Central and surface temperatures are labelled with $T_{\rm c}$ and $T_{\rm s}$.}}
\begin{align}\label{e:continuity}
 0 &= 2\pi r\, \partial_t \Sigma + \partial_r ( 2\pi r \Sigma v_r )\,,\\
\partial_r T &= 2\pi r\, \partial_t (\Sigma r^2 \Omega) +
 \partial_r (2\pi r v_r\Sigma r^2 \Omega \, )\,,
\label{e:momentum}
\end{align}
where the total torque $T=-T_\nu + T_{\rm d}$ is due to viscosity and the gravity of the secondary, given by
\begin{align}
T_\nu &= -2 \pi r^3 (\partial_r \Omega)\,\nu \Sigma \simeq 3\pi\, r^2
\Omega\, \nu \Sigma \,, \label{e:Tnu}\\
\partial_r T_{\rm d} &= 2\pi r \Lambda \Sigma \,.\label{e:Td}
\end{align}
Here $\Lambda$ is the torque per unit mass in the disk, approximately given by
\be\label{e:Lambda}
\Lambda \approx
\left\{
\begin{array}{c}
 - \frac{1}{2} f q^2 r^2\Omega^2  r^4/\Delta^4 \text{~~if~} r < r_{\SCO}\,,\\
 +\frac{1}{2} f q^2 r^2\Omega^2 r_{\SCO}^4/\Delta^4\text{~~if~} r > r_{\SCO}\,,
\end{array}
\right.
\ee
where
\begin{equation}\label{e:Delta}
 \Delta \equiv \max(|r - r_s|, H)
\end{equation}
$q \equiv m_{\SCO}/M_{\SMBH}$, $H\ll r$ is the scale-height of the
disk, and $f$ is a constant calibrated with simulations.  This
approximate formula for $\Lambda$, introduced by
\citet{2002ApJ...567L...9A}, accounts for the net contribution of all
Lindblad resonances as well as the torque cutoff within $r_{\SCO} \pm
H$ \citep{1980ApJ...241..425G,1997Icar..126..261W}, and guarantees
that the torque vanishes at $r\gg r_{\SCO}$.  Here
$f=(32/81\pi)[2K_0(2/3)+K_1(2/3)]^2=0.80$ outside the torque cutoff in
\citet{1980ApJ...241..425G}, $f=0.23\times (3/2\pi)=0.11$ in
\citet{1986ApJ...309..846L}, and $f=10^{-2}$ calibrated to match the
gap opening conditions in \citet{2002ApJ...567L...9A}.\footnote{
\citet{2010PhRvD..82l3011L} used Eq.~(\ref{e:Lambda}) with
$f=10^{-2}$.  \citet{2010MNRAS.407.2007C} adopted a torque model,
extrapolating Eq.~(18) of \citet{1980ApJ...241..425G} (with a modified
constant prefactor of $f=0.1\times (4/9\pi)\sim10^{-2}$), such that
their torque density approaches a constant at $r\gg r_{\SCO}$.
{ The linear perturbative analysis of \citet{1980ApJ...241..425G}
is not applicable if $q\gtrsim \alpha^{2}$ or $q \gtrsim (H/r)^3$
\citep{1987Icar...69..157M,1997Icar..126..261W,1996ApJS..105..181K}.} }
We adopt a conservative value
$f_{-2}\equiv f/10^{-2}\sim 1$ in our numerical calculations, but keep
the $f_{-2}$ terms general in all of our analytical formulas.  Note
that practically Eq.~(\ref{e:Delta}) assumes that the tidal torque
density ``saturates'' instead of having a true cutoff near the
secondary as long as the gas density is non-vanishing there
\citep{1993ApJ...419..155A}, which accounts for the effects of shocks
near the secondary
\citep{2001ApJ...552..793G,2011ApJ...741...57D,DM12}.\footnote{Recent
simulations \citep{2011ApJ...741...56D,2012ApJ...747...24R,DM12} have
shown that the original \citet{1980ApJ...241..425G} torque density is
correct close to the secondary, but the actual torque decreases in
amplitude and changes sign outside of $r_{\SCO} + 3 H$.  However the
relative contribution of these outer regions to the total torque is
negligible.}
\footnote{We do not account for relativistic corrections to the tidal
torque which are expected to be small at the separations in the
gas-driven regime $r_{\SCO}>100 M_{\SMBH}$
\citep{2011MNRAS.414.3198H,2011MNRAS.414.3212H}.}  However, this
prescription might be inaccurate for a high-mass secondary forming a
gap in the disk, where the tidal torques are due to spiral streams
passing near the secondary on horse-shoe orbits
\citep{2008ApJ...672...83M,2012ApJ...749..118S,2012arXiv1202.6063R,2012MNRAS.tmpL.436B,2012arXiv1203.5798P}.
We do not consider the torques inside the Hill radius, $|r-r_{\SCO}| <
r_{\rm H} \equiv (q/3)^{1/3} r_{\SCO}$, assuming that gas reaching
this region flows in across the secondary's orbit.  Outside this
region, we use Eq.~(\ref{e:Td}), assume that gravity is dominated by
$M_{\SMBH}$, and the orbital velocity is nearly Keplerian, $\Omega
\simeq M_{\SMBH}\R^{-3/2}$, where $\R= r/M_{\SMBH}$.

After some algebra \citep{2002apa..book.....F},
Eqs.~(\ref{e:continuity}--\ref{e:momentum}) simplify to
\begin{align}\label{e:gas}
 \dot{\Sigma}
 = -\frac{1}{2\pi r} \partial_r \left[ \frac{\partial_r T}{\partial_r (r^2 \Omega)} \right]\,,\quad
v_r
=\frac{\partial_r T}{2\pi r\, \Sigma \,\partial_r (r^2 \Omega)}\,.
\end{align}
The total mass flux across a ring of radius $r$ is defined as
\begin{equation}\label{e:mdot}
\dot{M}(r,t) \equiv -2\pi r \Sigma v_r = -\frac{ \partial_r T }{ \partial_r (r^2\Omega) }\,.
\end{equation}
Eq.~(\ref{e:gas}) along with the definition of the total torque $T$ in
Eqs.~(\ref{e:Tnu}--\ref{e:Lambda}) describes the evolution of the
axisymmetric disk surface density and radial velocity as a function of
radius and time.

The evolution of the secondary's orbital radius, $r_{\SCO}$, is driven
by the tidal torques of the gas and gravitational wave (GW)
losses. The angular momentum of the secondary is $L_{\SCO} = m_{\SCO}
r_s^2 \Omega_s$ so that
\begin{equation}\label{e:SCO}
\dot{L}_{\SCO} = \frac{1}{2} m_{\SCO} r_s \Omega_s v_{\SCO r} = -\int_0^{\infty} \partial_r T_{\rm d}\; \D r - T_{\GW}\,,
\end{equation}
where $-T_d$ is the recoil due to the torque exerted on the disk,
Eq.~(\ref{e:Td}), and with $\R_{\SCO} \equiv r_{\SCO}/M_{\SMBH}$, the
torque from the GWs is given by
\begin{equation}\label{e:Tgw}
T_{\rm GW} = \frac{32}{5} \frac{m_{\SCO}^2}{M_{\SMBH}} \R_{\SCO}^{-7/2}\,.
\end{equation}
Given $\nu(r,t)$ and $H(r,t)$, Eqs.~(\ref{e:gas}) and (\ref{e:SCO})
provide three equations for the three unknowns: $\Sigma(r,t)$,
$v_{r}(r,t)$, and $v_{\SCO r}(t)$.

We examine steady-state solutions to these equations where
$\dot{\Sigma}=0$ and $\D\dot M/\D r = 0$ so that $\dot{M}(r,t)\equiv
\dot{M}$ is a constant.\footnote{As stated above, we neglect the
accretion onto the secondary for simplicity (however, see
\citealt{1999ApJ...526.1001L}).}  Note that in general the disk need
not be in steady-state. However, in many cases the inflow rate of gas
may be much faster than the radial migration speed of the secondary
{ $|v_{\SCO r}|\ll |v_{r}|$. If this is satisfied in a wide range of radii up
to the outer edge of the disk,} then, the secondary is effectively
stationary in the azimuthally averaged picture, and the radial profile
of the disk might be expected to relax to a steady-state, independent
of the initial condition of the disk. We propose that the secondary
then migrates slowly through a sequence of quasi-steady-state
configurations of the disk with a fixed $\dot{M}(r,t)=\rm const$.
Then, Eq.~(\ref{e:mdot}) becomes
\begin{equation}\label{e:Tnu'}
 \partial_r T_{\nu} - \partial_r T_{\rm d} = \dot{M}\partial_r(r^2 \Omega)\,.
\end{equation}
This is a first-order ordinary differential equation for $T_{\nu}(r)$,
once $\partial_r T_{\rm d}(r)$ is specified for a specific disk model.

\subsection{Boundary conditions}\label{s:boundary}

We distinguish two types of inner boundary conditions corresponding to
whether { or not the perturbation is strong enough to
lead to a truncated disk with a wide hollow circular cavity. Here, ``wide'' means wider than the Hill radius (see below).}
\begin{enumerate}
 \item[I.] If $\Sigma(r) \neq 0$ all the way to the innermost stable
circular orbit $r_{\rm ISCO}$ of $M_{\SMBH}$ (i.e. the disk does not
have a cavity), we require a zero-torque boundary condition
\citep{1973blho.conf..343N,2010MNRAS.408..752P,2011MNRAS.410.1007T,2012arXiv1202.1530Z},
\begin{equation}\label{e:TnuISCO}
 T_{\nu}(r_{\rm ISCO})=0\,.
\end{equation}
Starting with this boundary condition, we obtain, among other
properties of the steady-state disk, the gas velocity profile
$v_r(r)$. As stated above, if this inflow velocity is much faster than
the migration velocity of the secondary over a large range of radii,
then one might expect that the disk approaches this steady-state
configuration, independent of the initial condition.  In the opposite
case, the steady-state assumption may be violated by the
time-dependent migration of the secondary.  As we will show, the
steady-state solution with a fixed $\dot{M}$ requires a large build-up
of gas outside the secondary for the viscosity to overcome the tidal
barrier of the secondary.  We refer to these solutions, in which
{ the disk is not truncated outside the secondary} as ``overflowing''.

 \item[II.] If the tidal torques dominate over the viscous torques
near the secondary, gas is expelled from the region near the secondary
and a { wide} gap forms.  Assuming that the characteristic radius $r_{\rm g}$,
where the tidal torque is exerted on the disk near the edge of the gap, tracks the
inward migration of the secondary with $r_{\rm g} = \lambda r_{\SCO}$
where $1< \lambda\lesssim 3$ is a constant,
we require that the gas velocity at that radius satisfy { \citep{1995MNRAS.277..758S,1999MNRAS.307...79I}}
\begin{equation}\label{e:vrdef}
 v_r(\lambda r_{\SCO})=\lambda v_{\SCO r}\,.
\end{equation}
Note that $\lambda$ is not specified by hand ab-initio; it is found
{ by assuming steady-state} in our solutions below.
This condition can be understood intuitively, since the secondary
cannot ``run away'' and leave the outer disk behind (if it did, it
would cease to be able to torque the disk and would have to slow
down).  Likewise, the gap edge cannot get closer to the secondary (if
it did, gas would pile-up and the gap would eventually
close). Although the disk is not in steady-state near its boundary, we
assume $\dot{M}(r) \approx \dot{M}$ at $r>r_{\rm g}$ (see discussion
{ in \S~\ref{s:middle:consistency}} below).

Based on Eq.~(\ref{e:mdot}) and (\ref{e:SCO}), Eq.~(\ref{e:vrdef}) is
equivalent to
\begin{equation}\label{e:vr1}
\int_{0}^{\infty} \partial_r T_{\rm d} \D r
= \frac{ m_{\SCO} r_{\SCO}^2 \Omega_{\SCO} \dot{M}}{4\pi r_{\rm g}^2 \Sigma(r_{\rm g})} - T_{\rm GW}\,.
 \end{equation}
\end{enumerate}

Note that { here and throughout the paper} by ``gap'' we refer to situations where the gas density
becomes { effectively} zero outside the secondary, such that the inflow of gas from
the outside pushes the secondary inward according to
(\ref{e:vrdef}). In these cases, we assume that inflow across the
orbit is insignificant, and in particular, we neglect torques from the
gas interior to the orbit. In our calculations, a gap is effectively a
hollow circular cavity in the disk, which is supported by the tidal
torques of the secondary. { However we emphasize that
we do not rule out the presence of a
local density decrement, resembling an annular gap,
with a significant mass flux across the gap.}

In practice, we attempt to find a solution with either of the above
two boundary conditions, and then check whether the solution is
self-consistent.  By construction, only one of the two boundary
conditions will lead to a self-consistent solution as confirmed below.

\subsection{Physical conditions in the disk}

Next we derive $H(r)$ and $\nu(r)$ which appear in the tidal and
viscous torques in Eqs.~(\ref{e:Tnu}--\ref{e:Lambda}).

\subsubsection{Vertical balance}\label{s:vertical}

Let us first derive the scale-height, $H$. If the vertical gravity is
dominated by $M_{\SMBH}$, (i.e. $|r-r_{\SCO}|>r_{\rm H}$), then in
vertical hydrostatic equilibrium $H=c_s/\Omega$ where
$c_s=\sqrt{p/\rho}$ is the local midplane sound speed and $p=p_{\rm
gas}+p_{\rm rad}$ is the pressure due to the gas and
radiation\footnote{Note that the gas is \emph{not} degenerate and is
\emph{not} isentropic, therefore the assumption of $p\propto
\rho^{5/3}$ or $\rho^{4/3}$ made in most numerical simulations of
accretion disks is inappropriate. In fact, $p\propto T^4 \propto
\dot{M}^{4}\rho^6$ for a radiation-pressure dominated standard
Shakura-Sunyaev disk with no secondary.}, $p_{\rm gas} = \rho k T_c /
(\mu m_p)$, $p_{\rm rad}=\frac{1}{3} {\rm a} T_{c}^4$, where $T_c$ is
the central temperature, $a=4\sigma/\C$ is the radiation constant,
$\sigma$ is the Stefan-Boltzmann constant, $m_p$ is the proton mass,
and $\mu=0.615$ is the mean particle mass in units of $m_p$.  Since
$\rho=\Sigma/(2H)$, $c_s^2 = 2Hp/\Sigma = 2c_s p/(\Sigma \Omega)$ so
that $c_s=2p/(\Sigma \Omega)$. The pressure can be expressed as
$p=p_{\rm rad}/(1-\beta)$, where $\beta=p_{\rm gas}/p$.  If photons
are transported to the surface by diffusion then the mean radiation
flux is
\begin{equation}\label{e:Fdef}
 F = \sigma T_{s}^4 = \frac{4}{3} \frac{\sigma T_{c}^4}{\tau} = \frac{8}{3} \frac{\sigma T_{c}^4}{\kappa \Sigma}
\end{equation}
Here $T_{s}$ is the surface temperature, $\tau = \kappa \Sigma/2$ is
the optical depth from the midplane to the surface, where
$\kappa=0.35\cm^2/{\rm g}$ is the opacity assumed to be dominated by
electron-scattering. We do not investigate changes caused by free-free
opacity at large radii for simplicity.  and neglect deviations from
blackbody radiation (see e.g. \citealt{2010ApJ...714..404T} for more a
detailed model).  Thus, $p_{\rm rad}=\frac{1}{2} \kappa \Sigma F/\C$,
so that $c_s=\kappa \C^{-1} F\, \Omega^{-1} (1-\beta)^{-1}$, and we
have
\begin{equation}\label{e:H}
H = \frac{c_s}{\Omega}= \frac{\kappa}{\C \Omega^2} \frac{F}{1-\beta}
\end{equation}
Note that Eq.~(\ref{e:H}) is valid in general for radiation flux
limited, geometrically thin disks, independent of the source of
dissipation and viscosity.\footnote{One possible source of
inconsistency is that convective vertical heat transport is
conventionally neglected here. This may be significant for optically
very thick, radiation pressure dominated disks (especially so-called
$\beta$ disks) with a large vertical temperature gradient
\citep{2011ApJ...733..110B}.  The heat transport in this regime may be
analogous to the convection zones of stars.}

\subsubsection{Viscosity}

In the standard Shakura-Sunyaev $\alpha$ and $\beta$--disk models, the
viscous stress tensor, $t_{ij}=\rho \nu \nabla_i v_{j}$ satisfies
$t_{r\phi}=-\frac{3}{2}\alpha \beta^b p$, where $b=0$ or $1$,
respectively, and $\alpha$ is a constant parameter
\citep{1973A&A....24..337S,1981ApJ...247...19S}, implying that
\begin{equation}\label{e:nu}
\nu = \alpha c_{s} H \beta^b = \alpha  \frac{\kappa^2}{\C^2 \Omega^3} \frac{ \beta^{b} F^2}{(1-\beta)^2}\,.
\end{equation}
In the second equality, we have substituted Eq.~(\ref{e:H}).

\subsubsection{Local thermal equilibrium}

We assume steady-state thermal equilibrium in which heat generated by
viscosity and the dissipation of the spiral density wave escapes the
optically thick disk in the vertical direction by photon diffusion.
The vertical radiation flux is $F = D_{\nu} + D_{\rm d}$. The viscous
dissipation rate per disk face element is
\begin{equation}\label{e:Dnu}
 D_{\nu} = \frac{(\partial_r \Omega) T_{\nu}}{4\pi r} = \frac{9}{8} \Omega^2 \nu \Sigma \,.
\end{equation}
We assume that the density waves generated by the tidal torque are
dissipated locally in the disk and turned into heat, yielding the rate
$D_{\rm d}$.  This is expected to be an adequate approximation based
on analytical arguments (\citealt{1980ApJ...241..425G}, Eq.~97
therein) and numerical studies
\citep{2011ApJ...741...57D,2012ApJ...747...24R,DM12}, especially in
the regime where the disk is strongly perturbed.

Following \citet{2001ApJ...552..793G} and
\citet{2009MNRAS.398.1392L}\footnote{We add a factor of 2 that appears
to be missing in \citet{2009MNRAS.398.1392L}; this enters because of
the two disk faces.},
\begin{equation}\label{e:Dd}
 D_{\rm d} =  \frac{(\Omega_{\SCO}-\Omega)\, \partial_r T_{\rm d} }{4\pi r}=\frac{1}{2}(\Omega_{\SCO} - \Omega) \Lambda \Sigma \,.
\end{equation}
The total vertical flux or total dissipation rate is
\begin{align}\label{e:F}
 F &=D_{\nu} + D_{\rm d}
 =
 \frac{9}{8} \Omega^2 \nu \Sigma + \frac{1}{2}(\Omega_{\SCO} - \Omega)  \Lambda \Sigma\,.
\end{align}
Using the above equations we derive $\Sigma$ and $T_{\rm c}$ for a
given $D_{\nu}$ and $F$ at each radius (see
Appendix~\ref{s:app:thermal}).

\subsubsection{Summary}
Combining the previous expressions, we obtain
\begin{align}\label{e:Sigma}
\Sigma &= \frac{8\,(\mu m_p/k)^{4/5}\sigma^{1/5}}{3^{9/5}\,
\alpha^{4/5} \kappa^{1/5}} \frac{\beta^{(1-b)4/5}}{\Omega^{4/5}}
\frac{D_{\nu}^{4/5}}{F^{1/5}}\,,\\
T_c &= \frac{(\mu m_p/k)^{1/5}\kappa^{1/5}}{3^{1/5}\, \alpha^{1/5}
\sigma^{1/5}}\frac{\beta^{(1-b)/5}}{\Omega^{1/5}}
F^{1/5}D_{\nu}^{1/5}\,,
\label{e:Tc}
\end{align}
where
\begin{align}\label{e:beta}
 \frac{\beta^{(1/2) + (b-1)/10}}{1-\beta} = \frac{\C [k /(\mu
m_p)]^{2/5} }{(3\,\alpha \sigma)^{1/10} \kappa^{9/10}} \Omega^{9/10}
\frac{D_{\nu}^{1/10}}{F^{9/10}}\,.
\end{align}
All other disk parameters can be derived from these relations. For
example, the scale-height $H$ and the quantity $\nu\Sigma$ that
determine the torque (Eq.~\ref{e:Tnu'}) are given by Eqs.~(\ref{e:H})
and (\ref{e:Dnu}).  In particular, the limiting cases for $H$ are
\begin{equation}
 H =
\left\{
 \begin{array}{ll}
 \kappa \C^{-1} \Omega^{-2}F & {~\rm if~~}\beta\ll 1\,,\\
 \sqrt{k/(\mu m_p)}\, \Omega\, T_c^{1/2} & {~\rm if~~}\beta\sim 1 \,.
\end{array}
\right.
\end{equation}
In the limit that the only source of heat is viscosity in a Keplerian
disk, $F=D_{\nu}=(3/8\pi) \dot{M}\Omega^2$, we recover the solution of
\citet{2003MNRAS.339..937G} up to a constant of order
unity.\footnote{We find a small difference in the density and
temperature normalization constants, due to
\citet{2003MNRAS.339..937G} neglecting a $4/3$ prefactor in the
vertical diffusion equation $F=\frac{4}{3} \sigma T_c^4/\tau$.}

More generally, Eqs.~(\ref{e:H}), (\ref{e:Sigma}), and (\ref{e:beta}),
along with the definition of $D_{\nu}$ and $F$ in Eqs.~(\ref{e:Dnu})
and (\ref{e:F}), and the angular momentum flow equation (\ref{e:Tnu'})
provide a closed set of equations for the stationary disk, valid
throughout the gas and radiation-pressure dominated regions for
$\alpha$ and $\beta$ disks.  The solution is self-consistent if for
all $r$, the disk is thin ($H<r$), the radiation flux is sub-Eddington
($L\sim 2\pi r^2 F<4\pi \C \G M_{\SMBH}/\kappa$), the radial accretion
velocity is subsonic ($v_r = \dot{M}/ 2\pi r\Sigma < c_{\rm s} =
H\Omega$), radial heat transport is negligible, the self-gravity of
the disk is negligible and the disk is stable against fragmentation
($Q=c_s \Omega/(\pi G \Sigma) \geq 1$), the disk is optically thick
($\tau = \kappa \Sigma/2 \geq 1$), and the boundary conditions are
satisfied (implying in particular that $v_{\SCO r}\ll v_r$ { across a wide range of radii for overflowing solutions},
see \S~\ref{s:boundary}).\footnote{ The model is
furthermore self-consistent only outside the secondary's Hill sphere
since the gravity of the secondary is accounted for as a perturbation
to the primary's gravitational field, and the equations are linearized
in the derivation of the torque formula. The tidal torque model is
nevertheless often interpolated to within this region, as well
\citep[e.g.][]{1980ApJ...241..425G,2002ApJ...567L...9A}. Here we avoid
this extrapolation by excising the region within the Hill radius from
our domain, assuming that gas entering this region flows across the
secondary orbit.}  We verify that these conditions are indeed
satisfied for the overflowing solutions below.

\section{Disk structure -- Numerical solutions}
\label{s:numerical}
First we generate numerical steady-state solutions for tidally and
viscously heated disks assuming that the migration rate is much
smaller than the radial accretion velocity in the disk. These
numerical solutions are useful to verify the detailed analytical
estimates presented in the following section.

We proceed along the following steps:
\begin{enumerate}
 \item\label{i:beta} Obtain the ratio of gas to total pressure,
$\beta=\beta(r,D_{\nu},F)$, by inverting Eq.~(\ref{e:beta}). A unique
solution is guaranteed by the intermediate value theorem, since the
left hand side is a monotonic function of $\beta$, mapping $0<\beta<1$
to all positive real numbers, while the right hand side is positive
and independent of $\beta$.
\item\label{i:nobeta} Substitute the solution for $\beta$ in
Eq.~(\ref{e:H}) and Eq.~(\ref{e:Sigma}) to obtain $H(r,D_{\nu},F)$ and
$\Sigma(r,D_{\nu},F)$.
 \item\label{i:F} Substitute $\beta$, $H$, and $\Sigma$ in the
definition of $F$, Eq.~(\ref{e:F}) to get an equation between $F$ and
$D_{\nu}$ for fixed $r$ and $r_{\SCO}$. Invert this relation to find
$F(r,r_{\SCO},D_{\nu})$. Similar to step~\ref{i:beta}, one can show
that the solution exists and is unique.
 \item\label{i:Td} Using Eqs.~(\ref{e:Dnu}--\ref{e:Dd}), obtain the
 function $\partial_r T_{\rm d} = g_{\rm d}(r, r_{\SCO},T_{\nu})$.
 \item\label{i:Tnu} Substitute into Eq.~(\ref{e:Tnu'}), to obtain an
expression $\partial_r T_{\nu} = g_{\nu}(r, r_{\SCO},T_{\nu})$ for a
fixed $\dot{M}$. Solve this differential equation for
$T_{\nu}(r,r_{\SCO})$.
 \item\label{i:final} Substituting back into $D_{\nu}$ and $F$,
Eqs.~(\ref{e:Dnu}) and (\ref{e:F}) and the formulas of
step~\ref{i:nobeta}, to get $\Sigma(r,r_{\SCO})$, $T_{\rm
c}(r,r_{\SCO})$, and $H(r,r_{\SCO})$.
\end{enumerate}
The complexity is related to the nonlinearities in steps~\ref{i:beta},
\ref{i:F}, and \ref{i:Tnu}.  Nevertheless, the solution exists and is
unique in steps~\ref{i:beta} and \ref{i:F}. However, step~\ref{i:Tnu}
is a boundary value problem of a nonlinear first-order differential
equation, which can have many solutions. We solve the differential
equation numerically upstream from an initial value $T_{\nu}(r_{\rm
ISCO})=0$. Without the secondary, the solution is simply $T_{\nu
0}(r)=\dot M (r^2 \Omega - r_{\rm ISCO}^2\Omega_{\rm ISCO})$, which
leads to the Shakura-Sunyaev disk.  If $q\ll 1$, then the secondary
creates a small dip in $T_{\nu}(r)$ in its neighborhood, where the
depth of the minimum increases with $q$. For larger $q$, $T_{\nu}(r)$
becomes very small positive approaching the secondary from downstream,
and the surface density approaches zero. In this regime, tidal heating
dominates over viscous heating, and $H>|r-r_{\SCO}|$, implying that
the pressure gradient shifts the torques out of resonance, and the
torque is suppressed according to Eq.~(\ref{e:Lambda}).  Since the
adopted torque model is valid only outside the secondary's Hill
radius, we stop the calculation at $r_{\SCO}-r_{\rm H}$, and restart
it at $ r_{\rm i} = r_{\SCO} + r_{\rm H}$ assuming that\footnote{The
tidal torque is monotonically increasing and decreasing, interior and
exterior the secondary, respectively. The solution is uniquely
determined by the initial value $T_{\nu}(r_{\ISCO})$ and $T_{\nu}(r_rm
i)$ in the two domains.  However, $r_{\rm i}$ can be arbitrary as long
as $r_{0}>r_{\SCO}$.}
 \begin{equation}\label{e:Tnur0}
T_{\nu}(r_{\SCO}-r_{\rm H}) \approx T_{\nu}(r_{\rm i})\,.
\end{equation}
This has a similar effect to smoothing the torque interior to the Hill radius as done previously in
\citet{1986ApJ...309..846L}, \citet{1995MNRAS.277..758S}, and \citet{2009MNRAS.398.1392L}.

The solution is approximately self-consistent if the migration rate is
slower than the radial gas velocity outside the secondary. However, if
this is not satisfied, { a cavity opens and the disk becomes truncated}.
In this case, we seek a different
solution in step~\ref{i:Tnu}, which satisfies the boundary condition
in Eq.~(\ref{e:vrdef}). This is possible by increasing $r_{\rm i}$ in
Eq.~(\ref{e:Tnur0}) where $T_{\nu}(r_{\rm i})\approx 0$, until
Eq.~(\ref{e:vr0}) is satisfied. Here $r_{\rm i}$ can be identified as
{ the truncation radius at } the inner edge of the disk.  We distinguish the characteristic { truncation or} gap
radius to reside at $r_{\rm g}$ where the tidal effect is exerted on
the disk, more specifically the boundary where the tidal torque
density becomes subdominant and use $r_{\rm g}$ in the boundary
condition, Eq.~(\ref{e:vr0}).\footnote{ In practice, we generate
solutions for many different $r_{\rm i}$.  We seek the radius $r_{\rm
g}$ at which the tidal torque cuts off in the numerical solution:
$T_{\rm d}(r_{\rm g})= 0.1\, T_{\rm d}(r_{\rm peak})$ where $r_{\rm g}
> r_{\rm peak}$ and $r_{\rm peak}$ is where $T_{\rm d}(r)$ attains its
maximum. We use this value as the gas velocity $v_{r}(r_{\rm g})$ in
Eq.~(\ref{e:vr0}).  We find that the gas velocity is nearly constant
in the neighborhood of $r_{\rm g}$ and the surface density is near its
peak, so the solution is insensitive to the details of this
convention.}  The surface density increases rapidly within $r_{\rm
i}<r\lesssim r_{\rm g}$ has a maximum and decreases thereafter. We
assume that the disk is truncated interior to $r_{\rm i}$ if a gap
forms with $r_{\rm i}>r_{\SCO}+r_{\rm H}$.

\section{Disk structure -- analytical solutions}\label{s:app:analytical}

Here we derive an analytical solution to the nonlinear equations in
\S~\ref{s:interaction}.  Such solutions can be derived asymptotically
far from the secondary or near the secondary, where either the tidal
torque or the viscous torque dominates, or where the angular momentum
flux is negligible.  We therefore distinguish the corresponding far,
middle, and near zones (see Figure~\ref{f:disk}). The \textit{far
zones} are well inside and well outside the secondary, where the
effects of the secondary are negligible.  The \textit{middle zone} is
the region outside the secondary where the tidal effects (i.e. torque
and heating) are locally negligible compared to the viscous effects,
but where the gas pile-up is significant and the disk profile is
modified.  The \textit{near zones} are just inside and just outside
the secondary's orbit, where the tidal effects of the secondary
dominate over the viscous effects. We restrict the near zone to
outside the Hill radius, where the adopted tidal torque formula is
valid.  In addition to providing a basic understanding of the disk
structure, the approximate analytical solutions allow us to infer the
migration rate of the secondary.

To keep track of the approximations and notations introduced for the
various zones below, we provide Table~\ref{t:approx} for convenience.
Note that the far/middle/near zones divide the disk into five radial
slices, and the asymptotic behavior further depends on whether
{ the disk becomes truncated forming a wide gap}
(in the middle zone) and whether the torque is saturated by
the condition on the radial distance from the secondary is $\delta
r\equiv r-r_{\SCO}<H$ (in the outer near-zone). Each row in the Table
corresponds to one of these disk regimes, discussed in a corresponding
subsection below, and shows which terms are relevant in
Eq.~(\ref{e:Tnu'}). The subdominant terms are marked with a ``0''.
The column with $T_{\nu}$ shows functions we introduced related to the
viscous torque, and $\partial_r T_{\rm d}$ shows the scaling of the
specific tidal torque in Eq.~(\ref{e:Lambda}).

\begin{table}
\begin{tabular}{lcccc}
\hline\hline
			&	\S\S			& $\dot{M}\partial_r(r^2\Omega)$ & $T_{\nu}$ 			& $\partial_r T_{\rm d}$\\	
\hline
 Far zone 	& \ref {s:far}		& \checkmark 		& $\varphi(r,r_{\ISCO})$				& 0	 \\
 Mid. {\it with gap}	 & \ref{s:app:gap}  & 0	 		& $T_{\rm bc}^{\rm mg}(r_{\SCO},r_{\rm g})$	 & 0	 \\
 Mid. {\it overflow}	 & \ref{s:middle}	& 0			& $T_{\rm bc}^{\rm mo}(r_{\SCO},r_{\rm i})$ 	 & 0	 \\
 Near {\it ext. uns.}	& \ref{s:nearout-un}  & 0	 		& $\zeta(r,r_{\SCO},r_{\rm i})$ 			& $r_{\SCO}^4/|\delta r|^4$	 \\
 Near {\it ext. sat.}	& \ref{s:nearout-sat} & 0			& $\psi(r,r_{\SCO},r_{\rm i})$				 & $r_{\SCO}^4/H^4$ \\
 Near interior	& \ref{s:nearin}	& \checkmark		& 0								& $-r^4/|\delta r|^4$\\
\hline\hline
\end{tabular}
\caption{\label{t:approx} Approximations and notations for the various
radial radial zones in the disk, used in \S~\ref{s:app:analytical}.  }
\end{table}

In the following we { mostly} focus on $\beta$--disks (i.e. $b=1$) and examine
both radiation and gas pressure dominated disks, but it is
straightforward to derive analogous formulas for $\alpha$--disks in
the same way.  We also note that in the radiation-pressure dominated
regime, the viscosity of $\alpha$--disks is larger by a factor of
$p_{\rm gas}/(p_{\rm gas}+p_{\rm rad})=\beta^{-1}$. This would
generally lead to stronger overflows for a smaller gas pile-up, and
the { cavity} would close for a wider range of parameters than we find below
for $\beta$ disks.

\subsection{Far and middle zones}

First we examine the region sufficiently far from the secondary,
either inside or outside of its orbit, where
\begin{equation}\label{e:Tnufarmid}
|\partial_r T_{\rm d}| \ll \partial_r T_{\nu}\approx \dot{M}\partial_r(r^2\Omega)\,.
\end{equation}
In this region, Eq.~(\ref{e:Tnu'}) can be integrated and substituted in (\ref{e:Dnu})
\begin{align}\label{e:Tnu1}
T_{\nu} &= \dot{M} r^2\Omega + T_{\rm bc}\,,\\
F &\approx D_{\nu} = \frac{3}{8\pi} \frac{\Omega}{r^{2}} T_{\nu} =
\frac{3}{8\pi} \left[ \dot{M} \Omega^2 + T_{\rm bc} \frac{\Omega}{r^{2}}\right]\,,
\label{e:F1}
\end{align}
where $T_{\rm bc}$ is an integration constant determined by the
boundary condition near the secondary. For a fixed $T_{\rm bc}$,
Eq.~(\ref{e:F1}) gives both $F$ and $D_{\nu}$, from which the surface density and
central temperature follow from Eqs.~(\ref{e:Sigma}--\ref{e:Tc}),
{
\begin{align}\label{e:Sigma_m}
\Sigma &= \frac{8^{2/5}\,(\mu m_p/k)^{4/5}\sigma^{1/5}}{(9\pi)^{3/5}\,
\alpha^{4/5} \kappa^{1/5}} \frac{\beta^{(1-b)4/5}}{\Omega^{4/5}}
\left[ \dot{M} \Omega^2 + T_{\rm bc} \frac{\Omega}{r^{2}}\right]^{3/5}\,,\\
T_c &= \frac{3^{1/5}(\mu m_p/k)^{1/5}\kappa^{1/5}}{(8\pi)^{2/5}\, \alpha^{1/5}
\sigma^{1/5}}\frac{\beta^{(1-b)/5}}{\Omega^{1/5}}
\left[ \dot{M} \Omega^2 + T_{\rm bc} \frac{\Omega}{r^{2}}\right]^{2/5}\,,
\label{e:Tc_m}
\end{align}
where
\begin{align}
 \frac{\beta^{(b+4)/10}}{1-\beta} = \frac{(8\pi)^{4/5} \C [k /(\mu
m_p)]^{2/5} }{3^{9/10}(\alpha \sigma)^{1/10} \kappa^{9/10}}
\frac{\Omega^{9/10}}{\left[ \dot{M} \Omega^2 + T_{\rm bc} \frac{\Omega}{r^{2}}\right]^{4/5}}.
\end{align}
Thus, solving the disk structure in these zones amounts to finding the torque
at the boundary, $T_{\rm bc}$.}

If $T_{\nu}(r_{\min})=0$ then Eq.~(\ref{e:Tnu'}) shows that,
\begin{equation}\label{e:c0def0}
 T_{\rm bc} = -  \dot{M}r_{\min}^2 \Omega(r_{\min}) +  \int_{r_{\min}}^{r} \partial_r T_{\rm d} \D r \,.
\end{equation}
In practice, $r_{\min}=r_{\ISCO}$ for a disk without a { cavity}, and it is
the inner edge of the disk if it has a { cavity}.  Depending on which term
dominates in Eq. (\ref{e:Tnu1}), we distinguish the far zone ($|T_{\rm
bc}| \ll \dot{M}r^2 \Omega$) and the middle zone ($|T_{\rm bc}| \gg
\dot{M}r^2 \Omega$). The far zone can be either well inside or far
outside the secondary's orbit, but the middle zone is always
outside. Well inside the secondary, the second term can be neglected
in Eq.~(\ref{e:c0def0}), and well outside of it, the second term
dominates and the integration domain can be extended to $\infty$.  In
both cases, $T_{\rm bc}$ is independent of $r$.

Eqs.~(\ref{e:Tnu1}) and (\ref{e:c0def0}) show, that in the region
outside the secondary, $T_{\rm bc}$ represents a {\it torque barrier}
due to the secondary's tidal effects.  This parameter can also be used
to obtain the migration rate of the secondary.  Indeed, combining
Eqs.~(\ref{e:SCO}) and (\ref{e:c0def0}) gives
\begin{equation}\label{e:vSCOc}
 v_{\SCO r} = -\frac{2 T_{\rm bc}}{m_{\SCO} r_{\SCO} \Omega_{\SCO}} - \frac{2 T_{\rm GW}}{m_{\SCO} r_{\SCO} \Omega_{\SCO}}\,.
\end{equation}

\subsubsection{Far zone -- unperturbed disk}\label{s:far}

Without the secondary Eq.~(\ref{e:TnuISCO}) implies that $T_{\rm
bc}=-\dot{M}r_{\ISCO}^2\Omega_{\ISCO}$.  Substituting into
Eqs.~(\ref{e:Tnu1}--\ref{e:F1}), gives $D_{\nu}$ and $F$.  Plugging
int Eqs.~(\ref{e:mdot}), (\ref{e:H}), and (\ref{e:Sigma}-\ref{e:Tc})
leads to the standard \citet{1973A&A....24..337S} solution
\begin{align}\label{e:Sigma0}
 \Sigma_0 &=
 4.7\times 10^5 \frac{\rm g}{{\rm cm}^2} \alpha_{-1}^{-4/5} \dot{m}_{-1}^{3/5} M_7^{1/5} r_{2}^{-3/5} \varphi^{3/5}\\
T_{c0} &= 5.4\times 10^5{\rm K}\, \alpha_{-1}^{-1/5}\dot{m}_{-1}^{2/5} M_7^{-1/5} r_{2}^{-9/10} \varphi^{2/5}\\
F_0 &=\frac{3}{8\pi}\dot{M}\Omega^2 =7.9\times10^{13}\frac{\rm erg}{{\rm s\, cm}^2} \dot{m}_{-1} M_{7}^{-1}  r_2^{-3} \varphi\\
 v_{r0} &= -3600 \frac{\rm cm}{\rm s} \alpha_{-1}^{4/5} \dot{m}_{-1}^{2/5} M_7^{-1/5} r_2^{-2/5} \varphi^{-3/5}\,,\label{e:vr0}\\
 H_0 &=
  \left\{
 \begin{array}{ll}
  1.5 M_{\SMBH}\,\dot{m}_{-1} \varphi &\mathrm{if}~\beta\ll 1\,,\\
  0.28 M_{\SMBH}\,\alpha^{-1/10}\dot{m}_{-1}^{1/5} M_7^{-1/10} r_{2}^{21/20}\varphi^{1/5} &\mathrm{if}~\beta\sim 1\,.
 \end{array}
 \right.\label{e:H0}
\end{align}
Here and below, the subscript $_0$ denotes quantities related to the
unperturbed disk, $\alpha_{-1}=\alpha/0.1$,
$\dot{m}_{-1}=(\dot{M}/\dot{M}_{\rm Edd})/0.1$, $\dot{M}_{\rm Edd}$ is
the Eddington accretion rate for $10\%$ radiative efficiency,
$q_{-3}=q/10^{-3}$, $M_7=M_{\SMBH}/10^7\,\Msun$, $r_{\SCO 2} =
r_{\SCO}/10^2 M_{\SMBH}$, and we introduced
\begin{equation}\label{e:varphi}
\varphi \equiv 1 - r_{\ISCO}^2\Omega_{\ISCO}/(r^2 \Omega) =1-(r_{\rm ISCO}/r)^{1/2}\,.
\end{equation}
Without the secondary, in the radiation pressure dominated regime
($\beta\ll 1$) the scale-height is approximately constant, and
increases approximately linearly further out where gas pressure
dominates ($\beta\sim 1$).

The viscous torque, for future reference:
\begin{equation}\label{e:Tnu0}
 T_{\nu 0} = \dot{M} r^2\Omega \varphi = 7.1\times 10^{47}{\rm erg}\, \dot{m} M_7^{2} r_2^{1/2} \varphi \,,
\end{equation}
where $r_2=r/10^2 M_{\SMBH}$. Sufficiently far from the secondary, the
disk is independent of the secondary and follows
Eqs.~(\ref{e:Sigma0}--\ref{e:H0}) with $\varphi\approx 1$.  However,
the disk structure depends on the rate at which gas is allowed to flow
in through $\dot{M}$.

\subsubsection{Middle zone}\label{s:middle}

Now let us consider the opposite limit, $T_{\rm bc} \gg \dot{M}r^2
\Omega$, where the steady-state perturbation to the torque is
significant.  In terms of the dimensionless torque barrier,
\begin{equation}\label{e:k}
    k=\frac{T_{\rm bc}}{\dot{M}r^2 \Omega} \equiv k_s \frac{r_{\SCO}^2 \Omega_{\SCO}}{r^2\Omega}\,,
\end{equation}
the formulae describing the unperturbed disk, Eq.~(\ref{e:Sigma0}--\ref{e:H0}),
get modified by replacing the boundary term with
\begin{align}\label{e:varphik}
\varphi \rightarrow 1+k\,.
\end{align}
The disk quantities change to
\begin{align}\label{e:Sigmam}
 \Sigma^{\rm m} &= \left(\frac{k+1}{\varphi}\right)^{3/5} \Sigma_0 \propto r^{-9/10}\\
 T_c^{\rm m} &= \left(\frac{k+1}{\varphi}\right)^{2/5} T_{c0} \propto r^{-11/10}\\
F^{\rm m} &= \left(\frac{k+1}{\varphi}\right) F_0 \propto r^{-7/2}\label{e:Fm}\\
v_r^{\rm m} &= \left(\frac{k+1}{\varphi}\right)^{-3/5} v_{r0} \propto r^{-1/10}\\
 H^{\rm m} &=
 \left\{
 \begin{array}{ll}
 (k+1)H_0/\varphi  \propto r^{-1/2} &\mathrm{if}~\beta\ll 1\,,\\
 ((k+1)/\varphi)^{1/5}H_0  \propto r^{19/20} &\mathrm{if}~\beta\sim 1\,.
 \end{array}
\right.\label{e:Hm}
\end{align}
and the migration rate follows from Eq.~(\ref{e:vSCOc})
\begin{equation}\label{e:vSCOrm}
v_{\SCO r}\approx -\frac{2T_{\rm bc}}{m_{\SCO} r_{\SCO} \Omega_{\SCO}}
= -2 k_{\SCO} \frac{ \dot{M} r_{\SCO}}{m_{\SCO}}
\end{equation}
where we have assumed $T_{\rm GW}\ll T_{\rm bc}$.  Here and below, the
superscript ``m'' labels the middle zone.  Note that { the dimensionless
angular momentum flux} $k$ can be
interpreted as a \textit{brightening factor} in the middle zone
relative to the unperturbed disk; $k_{\SCO}$ is representative of the
maximum brightening, if $k(r)$ is extrapolated to $r_{\SCO}$. In
practice, the maximum brightening is even larger than $k_{\SCO}$ in
the near zone due to tidal heating (see \S~\ref{s:nearout-middle}
below).

{
The disk is modified within a radial range where the
dimensionless angular momentum flux satisfies $k>1$. This sets the outer boundary
$r^{\rm m}_{\rm f}$ of the middle zone, where the disk transitions to the far zone. From Eq.~(\ref{e:k}),
\begin{align}\label{e:rt}
r^{\rm m}_{\rm f} &= \frac{T_{\rm bc}^2}{\G M_{\SMBH} \dot{M}^2} =
k_{\SCO}^2 r_{\SCO}\,.
\end{align}
}

Eqs.~(\ref{e:Sigmam}--\ref{e:Hm}) represent a disk with negligible
inflow of angular momentum but an inner boundary condition with a
large viscous torque, corresponding to the torque barrier.  Such
solutions are often (somewhat misleadingly) referred to as a {\it
decretion disk} \citep{1991MNRAS.248..754P,2009MNRAS.398.1392L}.  To
avoid confusion, we emphasize that there is accretion (i.e. inflow) in
this region, too, with a fixed $\dot{M}$.  However, the radial
accretion velocity is greatly reduced, while the surface density,
temperature, and scale-height are all greatly increased, relative to
an accretion disk around a single compact object.

So far in this subsection, we have derived a solution for an arbitrary
torque barrier or $k$, without specifying its value. In general, $k$
is given by Eq.~(\ref{e:c0def0}), which depends on the tidal torque in
the near zone. Thus, to complete the derivation of the disk structure
in the middle zone, we are first required to obtain the disk structure
in the near zone (which we will do in \S~\ref{s:near} below). However,
in the case of the steady-state { cavity}, the particular form of the
boundary condition allows us to directly infer $k$, independently of
the near zone, up to a factor $\lambda$ of order unity, which we show
next.

\subsubsection{Middle zone -- steady-state disk with a cavity}\label{s:app:gap}

When the tidal torque is sufficiently strong to clear a gap so that
the secondary and the nearby gas move with a similar velocity,
$T_{\nu}\sim T_{\rm bc}$ can be substantial over a large range of
radii.  From Eq.~(\ref{e:vrdef}--\ref{e:vr1}), this requires
\begin{equation}\label{e:c0def1}
T_{\nu}^{\rm mg} = T_{\rm bc}^{\rm mg} = \int_0^{\infty} \partial_r T_{\rm d} \,\D r
=
\frac{ m_{\SCO} r_{\SCO}^2 \Omega_{\SCO} \dot{M}}{4\pi r_{\rm g}^2 \Sigma(r_{\rm g})}\,.
\end{equation}
Here and below, the superscript $^g$ refers to solutions with a gap,
and $r_{\rm g}=\lambda r_{\SCO}$ is the outer radius of the gap.  For
this value of $T_{\rm bc}$, Eq.~(\ref{e:F1}) gives $F$ and $D_{\nu}$, and
$\Sigma$ follows from (\ref{e:Sigma}). However, since the right hand
side (RHS) of Eq.~(\ref{e:c0def1}) depends on $\Sigma$ itself, this
gives an algebraic equation for $T_{\rm bc}$. The solution is
\begin{align}\label{e:c0}
T_{\rm bc}^{\rm mg}
&=\frac{3^{3/4} \alpha^{1/2} \kappa^{1/8}}{4 \pi^{1/4}(\mu m_p/k)^{1/2}\sigma^{1/8}}
m_{\SCO}^{5/8} \dot{M}^{5/8} \frac{\Omega_{\SCO}^{3/4} r_{\SCO}^{3/4}}{\lambda^{11/16}}\\\nonumber
&=1.6\times 10^{49}\,{\rm erg}\, \alpha_{-1}^{1/2} \dot{m}_{-1}^{5/8} \lambda^{-11/16} q_{-3}^{5/8} M_7^{5/4} r_{\SCO 2}^{-3/8}\,.
\end{align}
Note that this is independent of the tidal torque model (i.e. the
$\Lambda$ or $f$ in Eq.~(\ref{e:Lambda})), since here the tidal torque
is set by the boundary condition of the gap. This solution breaks
down, and becomes tidal torque dependent, if the gap closes, which we
discuss in \S~\ref{s:near} below.

The { dimensionless angular momentum flux} from Eq.~(\ref{e:k}) is
\begin{align}\label{e:kmg}
k^{\rm mg} = 23\, \alpha_{-1}^{1/2} \dot{m}_{0.1}^{-3/8} M_7^{-3/4} q_{-3}^{5/8} \lambda^{-19/16} r_{\SCO 2}^{-7/8}
\left(\frac{r}{\lambda r_{\SCO}}\right)^{-1/2}\,.
\end{align}
In particular, near the secondary $k^{\rm mg}(r_{\rm
\SCO})=m_{\SCO}/(4\pi r_{\rm g}^2 \Sigma_{\rm g})$ is the ratio of
secondary mass to the accumulated local gas mass.\footnote{Here
$k^{\rm mg}(r_{\rm \SCO}) = B^{-5/8}$ using the Syer-Clarke parameter
$B=m_{\SCO}/(4\pi r_{\SCO}^2 \Sigma_{\SCO 0})$.}  The only free
parameter in this zone is $\lambda$, which we determine explicitly in
\S~\ref{s:nearout-gap} below.

In the range  $r_{\SCO}\ll r \ll r^{\rm mg}_{\rm f}$, $k^{\rm mg}\gg 1$ and Eqs.~(\ref{e:Sigmam}--\ref{e:Hm}) give
\begin{align}\label{e:Sigma2b}
\Sigma^{\rm mg}
&= 3.1\times 10^6 \frac{\rm g}{\rm cm^2}
\frac{\dot{m}_{-1}^{3/8}}{\alpha_{-1}^{1/2}}  \frac{q_{-3}^{3/8}}{M_7^{1/4}}
\lambda^{-33/80} r_{\SCO 2}^{-9/40} r_{2}^{-9/10}\,,\\
T_c^{\rm mg}
&= 1.9\times 10^6\,{\rm K}\,
\dot{m}_{-1}^{1/4}  \frac{q_{-3}^{1/4}}{M_7^{1/2}}
\lambda^{-11/40} r_{\SCO 2}^{-3/20} r_{2}^{-11/10}\,,\\
v_{r}^{\rm mg} &= \lambda\, v_{\SCO r} \left(\frac{r}{\lambda r_{\SCO}}\right)^{-1/10}\\
H^{\rm mg} &=
 \left\{
 \begin{array}{l}
35M_{\SMBH}\,\alpha_{-1}^{1/2}\dot{m}_{-1}^{5/8} M_7^{-3/4} \lambda^{-19/16} q_{-3}^{5/8} r_{\SCO 2}^{-7/8} \\
\hspace{15pt}\times(r/\lambda r_{\SCO})^{-1/2}~~{\rm if~}\beta\ll 1\,,\\
0.53M_{\SMBH}\,\dot{m}_{-1}^{1/8} M_7^{-1/4} \lambda^{-11/16} q_{-3}^{1/8} r_{\SCO 2}^{-3/40}
r_{2}^{19/20}\\
\hspace{15pt}{\rm if~}\beta\sim 1\,.
 \end{array}
 \right.\label{e:H2b}
\end{align}
where $\Sigma(r) \sim 0$ at $r\leq \lambda r_{\SCO}$.  Outside $r\gg
r^{\rm mg}_{\rm f}$, $k^{\rm mg}\approx 0$, and the disk approaches
the unperturbed solution given by
Eqs.~(\ref{e:Sigma0}--\ref{e:H0}). In the transition zone, between the
middle and far zones, $r\sim r^{\rm mg}_{\rm f}$, one needs to use
Eqs.~(\ref{e:Sigmam}--\ref{e:Hm}) with $k=k^{\rm mg}$
(Eq.~\ref{e:kmg}).

When a cavity is present, the migration speed of the secondary follows from
Eqs.~(\ref{e:vSCOc}) and (\ref{e:c0}):
\begin{align}
v_{\SCO r}^{\rm g} &= -550\frac{\rm cm}{\rm s} \alpha_{-1}^{1/2}
\dot{m}_{-1}^{5/8} M_7^{1/4} q_{-3}^{-3/8} \lambda^{-11/16} r_{\SCO
  2}^{1/8}.
\label{e:vSCOr_mg}
\end{align}
This expression is consistent with the secondary dominated Type-II
migration rate of \citet{1995MNRAS.277..758S} who assumed $\lambda=1$.
Note that the migration speed is slower than the gas inflow velocity
without the secondary, $|v_{\SCO r}^{\rm mg}|<|v_{r0}(r_{\SCO})|$ in
Eq.~(\ref{e:vr0}).  This is referred to as disk-dominated Type-II
migration, which is appropriate if the secondary mass is smaller than
the unperturbed local disk mass $m_{\SCO} \leq 4\pi r_{\rm g}^2
\Sigma_0(r_{\rm g})$ (or equivalently $k^{\rm mg}\geq 1$), but large
enough to open a gap.

It is interesting to note that the structure of the middle zone does
not depend explicitly on the tidal torque model, $\partial_r T_{\rm
d}$ (in particular $\Lambda$ or the $f_2$ parameter in
Eq.~\ref{e:Lambda}); the dependence is implicit and arises only by
fixing the value of $\lambda$.  Physically, while the tidal torques
are negligible in this region, the effects of the tidal torques are
still communicated to the region by setting an effective
hydrodynamical boundary condition.  We determine $\lambda$ in
\S~\ref{s:nearout-gap} below and find that, in fact, it only weakly
depends on $\partial_r T_{\rm d}$.

\subsubsection{ { Consistency of steady--state} }\label{s:middle:consistency}

{
A basic assumption of our model is that the radial structure of the disk is in a quasi steady--state
as the secondary migrates slowly inwards. }
To check the consistency of these steady-state solutions, we must
verify that the implicit time-dependence in the surface density
profile through $r_{\SCO}(t)$ does not violate the continuity
equation~(\ref{e:continuity}) significantly, so that
\begin{equation}
 \partial_r \dot{M} = -\partial_r (2 \pi r \Sigma v_{r}) = -2 \pi r \partial_{t} \Sigma \stackrel{?}{=} 0
\end{equation}
Integrating over radius, the relative error in the accretion rate
\begin{align}
\frac{\int 2\pi r \dot{\Sigma}^{\rm mg} \D r}{\dot{M}}\ =
\frac{\int 2\pi r v_{\SCO r}^{\rm mg}\partial_{r_{\SCO}} \Sigma^{\rm mg}\,  \D r}{2\pi r v_r^{\rm mg} \Sigma^{\rm mg} }
= \frac{9}{44} \left(\frac{r}{\lambda r_{\SCO}}\right)^{11/10}
\end{align}
which is $\sim 20\%$ near the gap edge. The error in $T_{\rm bc}$
based on Eq.~(\ref{e:c0}) is $\sim 13\%$. However, the error in the
accretion rate exceeds unity at large radii, outside $4.2 \lambda
r_{\SCO}$. The steady-state assumption breaks down because as the
secondary migrates inward, the steady-state gas density near the edge
of the gap continuously increases with time. If $\dot M$ is fixed near
the gap edge to be a constant fraction of the Eddington value, the
true accretion rate $\dot{M}(r)$ at larger radii must be larger, to
supply material for the increasing gas density. Conversely, if
$\dot{M}(r)$ is fixed at large radii, then it becomes smaller
approaching the gap edge.  Such non-steady-state solutions have been
derived by \citet{1991MNRAS.248..754P} and \citet{1999MNRAS.307...79I}
by solving the nonlinear diffusion equation (\ref{e:gas}) {\it for a
fixed outer boundary condition}, assuming that the viscosity can be
expressed as $\nu = k \Sigma^a r^b$ where $k$, $a$, and $b$ are
constants.  In particular, \citet{1999MNRAS.307...79I} derived a
non-steady, but self-similar solution.  In that solution, the
migration is slower, and the { angular momentum flux} is lower, compared to
{ the \citet{1995MNRAS.277..758S} steady-state solutions with
 a fixed $\dot{M}$} for the same binary and
disk parameters \citep{1995MNRAS.277..758S}.  { The quasi-steady} migration rate and
brightening factors  { for a truncated disk with $\lambda>1$ are} intermediate
between the \citet{1995MNRAS.277..758S} and
\citet{1999MNRAS.307...79I} solutions.

{
The steady-state condition is typically \emph{not} violated in the overflowing solution
over a wide radial range. For global-steady state, a necessary condition is
\begin{equation}
\left|\frac{\int_0^{r^{\rm m}_{\rm f}} 2\pi r \dot{\Sigma}^{\rm m} \D r}{\dot{M}}\right| =
 \left|\frac{\gamma_{\Sigma {\SCO}}}{2+\gamma_{\Sigma r}}\right|\,  \,k_{\SCO}^{19/5} \frac{4\pi r_{\SCO}^2 \Sigma_0(r_{\SCO})}{m_{\SCO}} \ll 1
\end{equation}
where we have used Eqs.~(\ref{e:Sigma_m}), (\ref{e:vSCOrm}), and (\ref{e:rt})
and defined $\gamma_{\Sigma {\SCO}} = \partial \ln \Sigma^{\rm m}/\partial \ln r_{\SCO}$
and
$\gamma_{\Sigma r} = \partial \ln \Sigma^{\rm m}/\partial \ln r$.
This sets a maximum limit for the dimensionless angular momentum flux
$k_{\SCO}$. For a $\beta$-disk, this implies
\begin{equation}\label{e:kmax}
 k_{\SCO \max} = 3.2\, |\gamma_{\Sigma {\SCO}}|^{5/19} \alpha_{-1}^{4/19} \dot{m}_{-1}^{-3/19} M_7^{6/19} q_{-3}^{5/19} r_{\SCO 2}^{-7/19}\,,
\end{equation}
and $k_{\SCO \max}$ is larger (i.e. less restrictive)
for radiation pressure dominated $\alpha$-disks.

An important qualitative difference between the overflowing model presented here and the
\citet{1995MNRAS.277..758S} model for a truncated disk is that
$\gamma_{\Sigma {\SCO}}>0$ for the former as we show below.
In contrast, in the overflowing case, the excess surface density
and the dimensionless angular momentum flux  $k_{\SCO}$ in the middle zone
both gradually decrease during the inward migration of the secondary.
Thus, the excess surface density diffuses radially outwards.
If $k_{\SCO}< k_{\SCO \max}$ then the diffusion is
sufficiently fast to reach a global quasi-steady-state
throughout the middle zone. If this is not satisified, then the outer parts
of the middle zone cannot respond as quickly as the object moves inwards
and the structure of the disk in these regions will depend on its previous history.
However, since the viscous timescale is always much smaller than the migration
timescale in at least the inner parts of the middle zone, the local disk structure
of the overflowing solution might approach an approximate steady-state
there with a constant $\dot{M}$ even if $k_{\SCO} \gtrsim k_{\SCO \max}$.
The migration rate of the secondary depends on the near-zone of the disk, which
is expected to remain insensitive to perturbations
in the outer parts of the middle zone in an overflowing disk.\footnote{{ This is different from a
transient truncated circumbinary disk where the migration velocity is comparable to the local gas accretion velocity,
which can exhibit hysteresis throughout the middle zone \citep{2012arXiv1205.5017R}.}}
We leave a detailed investigation of the time dependent overflowing solutions to future work \citep[][in preparation]{Munier}.
}

\subsection{Near zone}\label{s:near}

Now let us consider the regions near the secondary where the tidal
torque and heating are important.  We discuss steady-state solutions
inside and outside of the secondary's orbit, in turn, without and then with a { circular cavity}.
Deriving the physical properties of the disk in
this region is useful to provide an estimate of the torque barrier,
$T_{\rm bc}$, at the interface between the near zone and the middle
zone. As explained previously, the torque barrier sets the overall
scale of the physical properties in the middle zone, as well as the
migration rate of the secondary.  We therefore first compute the value
of the torque barrier for an overflowing disk, as well as for a disk
with a { wide} gap.  In the latter case, we then compare the value with the
torque barrier in the middle zone derived above (Eq.~\ref{e:c0}). By
equating the two, we can estimate the gap size (i.e. $\lambda$), and
obtain the conditions for gap opening and closing.

\subsubsection{Inside the secondary orbit}\label{s:nearin}

Consider the region just downstream the secondary, outside the torque
cutoff $\Delta = r_{\SCO}-r > H$ in Eq.~(\ref{e:Delta}), assuming a
steady-state overflow (i.e. { no hollow circular cavity}).
Based on Eq.~(\ref{e:Tnu'}), the viscous torque decreases in the
vicinity of the secondary and for a sufficiently large secondary mass,
the angular momentum exchange is dominated by the tidal torque.  In
this regime,
\begin{equation}\label{e:dTnu-ni}
|\partial_r T_{\nu}| \ll |\partial_r T_{\rm d}| = 2\pi r |\Lambda|
 \Sigma \approx \dot{M}\partial_r(r^2 \Omega)
\end{equation}
implying that
\begin{equation}\label{e:Sigmain}
  \Sigma^{\rm ni} =  \frac{\dot{M}}{4\pi} \frac{\Omega}{|\Lambda|}
=\frac{\dot{M}}{2\pi f q^2 } \frac{1}{r^2\Omega}\frac{\Delta^4}{ r^4}
\end{equation}
for a Keplerian disk (here and below, the superscript $^{ni}$ refers
to the solutions in the inner near zone).  Eq.~(\ref{e:F}) shows that
\begin{align}\label{e:Fin}
 F^{\rm ni} &\approx D_{\rm d} = \frac{1}{2}(\Omega_{\SCO} - \Omega) \Lambda \Sigma =
\frac{\dot{M}}{8\pi}\Omega(\Omega - \Omega_{\SCO}) \rightarrow \frac{F_0}{2} \frac{\Delta}{r}\,.
\end{align}
The asymptotic limit corresponds to $\Delta\ll r$.  From
Eqs.~(\ref{e:mdot}) and (\ref{e:H})
\begin{align}
 v_r^{\rm ni} & = \frac{2 \Lambda}{ r  \Omega} = f q^2 \frac{r^5 \Omega}{\Delta^4}\\\label{e:Hni0}
 H^{\rm ni} &=\frac{\kappa \dot{M}}{8 \pi c}\frac{\Omega - \Omega_{\SCO}}{\Omega}
\rightarrow \frac{H_0}{2} \frac{\Delta}{r} ~~\mathrm{if}~\beta\sim 1
\end{align}
The latter equation shows that the secondary makes the disk thinner
downstream if the disk is radiation pressure dominated.  Combining
Eqs.~(\ref{e:Sigmain}--\ref{e:Fin}) with (\ref{e:Sigma}), gives
$D_{\nu}^{\rm ni}$.  The viscous torque then follows from
Eq.~(\ref{e:Dnu}). To first beyond leading order, for $b=1$,
\begin{align}
 T_{\nu}^{\rm ni} &= 2.4\times 10^{50}{\rm erg}\,\alpha_{-1} \dot{m}_{-1}^{3/2} f_{-2}^{-5/4} M_7^{7/4} q_{-3}^{-5/2} \nonumber\\
&\quad\times r_{2}^{5/8} \left(\frac{r}{r_{\SCO}}\right)^{5/16} \left(\frac{\Delta}{r}\right)^{21/4}\,,
\label{e:Tnuin}
\end{align}
$T_{\nu}^{\rm ni}$ exhibits a sharp cutoff near the secondary.

We can verify that the working assumptions hold in this
region. Eq.~(\ref{e:Hni0}) shows that $\Delta > H$ holds for all
$\Delta$, since the unperturbed disk is thin, $H_0<r$.  Since
$D_{\nu}^{\rm ni}\propto T_{\nu}^{\rm ni} \propto \Delta^{21/4}$ which
implies that $D_{\nu}^{\rm ni}\ll F^{\rm ni}$ is indeed satisfied for
sufficiently small $\Delta$.  Coincidentally, the assumption in
Eq.~(\ref{e:dTnu-ni}) is satisfied within a distance $\Delta_{\rm ni}$
from the secondary, where
\begin{equation}
 \frac{\Delta_{\rm ni}}{r_{\SCO}} = \frac{x_{\rm ni}}{1+\frac{841}{714}x_{\rm ni}}
\end{equation}
and
\begin{equation}
 x_{\rm ni} = 0.1\, \alpha_{-1}^{-4/17}  \dot{m}^{-2/17}  M_7^{1/17} f_{-2}^{5/17}q_{-3}^{10/17} r_{\SCO 2}^{-1/34}\,.
\end{equation}
The disk parameters in the region $r_{\SCO}- \Delta^{\rm ni}\lesssim r
\lesssim r_{\SCO}-r_{\rm H}$ are
\begin{align}
 \Sigma^{\rm ni} &= 5.7\times 10^{7}\frac{\rm g}{\rm cm^2} f_{-2}^{-1} \dot{m}_{-1} q_{-3}^{-2} r_{2}^{-1/2} (\Delta/r)^{4}\,,\\
 T_{c}^{\rm ni} &=  1.5\times 10^6{\rm K}\, f_{-2}^{-1/4} \dot{m}_{-1}^{1/2} M_7^{-1/4} q_{-3}^{-1/2} r_{2}^{-7/8}
\nonumber\\&\quad\times
\left(\frac{r}{r_{\SCO}}\right)^{5/16}
\left(\frac{\Delta}{r}\right)^{5/4},\\
F^{\rm ni} &= 3.9\times10^{13}\frac{\rm erg}{{\rm s\, cm}^2} \dot{m}_{-1} M_{7}^{-1}  r_2^{-3} \left(\frac{r}{r_{\SCO}}\right)^{5/4} \frac{\Delta}{r}\,,\\
v_r^{\rm ni} &= 30 \frac{\rm cm}{\rm s} f_{-2} q_{-3}^{2} r_2^{-1/2} (\Delta/r)^{-4}\,,\\
H^{\rm ni} &=
 \left\{
 \begin{array}{l}
 0.75 M_{\SMBH}\, \dot{m}_{-1} (r/r_{\SCO})^{5/4} (\Delta/r) \quad\mathrm{if}~\beta\ll 1\,,\\
0.36 M_{\SMBH}f_{-2}^{-1/8} \dot{m}_{-1}^{1/4} M_7^{-1/8} q_{-3}^{-1/4}r_2^{17/16} \\
\quad \times  (r/r_{\SCO})^{5/32} (\Delta/r)^{5/8} \quad\mathrm{if}~\beta\sim 1\,.
 \end{array}
\right.
\end{align}
During the inward migration of the secondary, $r_{\SCO}$ decreases,
and the surface density evolves in a self-similar way.  The surface
density, midplane temperature, surface brightness, and scale-height
all decrease significantly near the secondary with large $q_{-3}$. The
radial flow velocity becomes very large in the close vicinity, and the
flow may become advection dominated there.  However, we do not
extrapolate this solution inside the Hill radius of the secondary
because the torque model is invalid there.

It is remarkable that for a fixed $\dot{M}$, the fractional
perturbation to the surface brightness and the
radiation-pressure-dominated scale-height are universal in this
region, independent of the binary and disk parameters. This property
is general for an arbitrary torque or viscosity model in radiatively
efficient steady-state disks.  The surface density in this regime is
also independent of the viscosity model but it is sensitive to the
torque model: it is set to ensure that the tidal torque matches the
angular momentum flow associated with $\dot{M}$.  The original value
of the surface density is suppressed by a factor proportional to
$q^{-2} \Delta$.  These solutions are valid only for disks with
relatively large secondary masses, such that $\Delta_{\rm ni}>r_{\rm
H}$, but in which there is gas inflow across the secondary orbit.

\subsubsection{Outside the secondary -- unsaturated torque}\label{s:nearout-un}

Next consider the region just outside the secondary.  Here we examine
the case where the secondary is massive enough for the tidal torques
to be important. After a significant amount of gas pile--up the
viscous torque eventually counteracts the tidal torque and creates a
stationary inflow. In this regime the tidal and viscous torques
counteract one another and both greatly exceed the momentum flux in
Eq.~(\ref{e:Tnu'}) such that
\begin{equation}\label{e:dTnuout}
  \dot{M} \partial_r (r^2\Omega) \ll \partial_r T_{\nu} \approx
  \partial_r T_{\rm d} = 2\pi r \Lambda \Sigma\,.
\end{equation}
The tidal heating rate is much larger than the viscous heating rate of
a disk without a satellite, making the disk much hotter and
thicker. Let us assume $D_{\rm d}\gg D_{\nu}$ here, Eq.~(\ref{e:F})
implying that
\begin{equation}\label{e:Foutdef}
F \approx D_{\rm d} = \frac{1}{2} (\Omega_{\SCO} - \Omega) \Lambda
\Sigma \,.
\end{equation}
In this subsection we examine the case where the torque is not
saturated, $H(r) < r - r_{\SCO}$, so that $\Delta = r-r_{\SCO}$ in
Eq.~(\ref{e:Delta}). This is most relevant for relatively small mass
ratios, (i.e. typically $q\lesssim 10^{-3}$, see Eq.~\ref{e:qneu}
below and Paper~II), where the banking up of the stream in this region
is modest.  Substitute Eq.~(\ref{e:Sigma}) for $\Sigma$ with $b=1$,
and use Eq.~(\ref{e:Dnu}),
\begin{equation}\label{e:Foutunsat2}
F = \left(\frac{3}{8\pi}\right)^{4/5} \frac{a_{\Sigma}}{2}
(\Omega_{\SCO} - \Omega) \Lambda r^{-2/5}
\frac{T_{\nu}^{4/5}}{F^{1/5}}
\end{equation}
where $a_{\Sigma}$ is the constant coefficient in Eq.~(\ref{e:Sigma}).
Solve this for $F$ and plug back into Eq.~(\ref{e:Foutdef})
\begin{equation}\label{e:Tnuoutunsat2}
\frac{\partial_r T_{\nu}}{T_{\nu}^{2/3}} = \frac{2^{13/19}
\pi^{13/15}}{3^{2/15} a_{\Sigma}^{1/6}} \frac{ r \Lambda^{5/6}}{
(\Omega_{\SCO}-\Omega)^{1/6} }
\end{equation}
Integrating both sides between $r_{\rm i}$, the inner edge of this
region (see further below for a discussion of the value of $r_{\rm
i}$) and $r$,
\begin{equation}\label{e:Tnuoutunsat3}
T_{\nu}^{1/3}(r) - T_{\nu}^{1/3}(r_{\rm i}) = \frac{2^{13/19}
\pi^{13/15}}{3^{17/15} a_{\Sigma}^{1/6}} \int_{r_{\rm i}}^{r} \frac{ r
\Lambda^{5/6}}{ (\Omega_{\SCO}-\Omega)^{1/6} } \D r\,.
\end{equation}
We are most interested in the case where the tidal torques increase
$T_{\nu}$ substantially in this region.  If so, we approximate
$T_{\nu}(r_{\rm i})=0$.  Substituting Eq.~(\ref{e:Lambda}), and
rearranging gives
\begin{equation}\label{e:Tnuoutunsat4}
T_{\nu}^{\rm neu} = 3.5\times 10^{40}{\rm erg}\, \alpha_{-1}^{-2} M_7^{5/2} f_{-2}^{5/2} q_{-3}^5 r_{\SCO 2}^{1/4}
\zeta^{3}(r,r_{\SCO}, r_{\rm i})
\end{equation}
where the superscript $^{neu}$ refers to the case of unsaturated
torque in the external near zone, and
\begin{align}\label{e:zeta0}
 \zeta &\equiv \int_{r_{\rm i}/r_{\SCO}}^{r/r_{\SCO}} x^{-7/6} \left(1-x^{-3/2}\right)^{-1/6}(x - 1)^{-10/3}\,\D x\\\nonumber
&\approx \left(\frac{2}{3}\right)^{1/6}\frac{2}{5}
\frac{(r_{\rm i}/r_{\SCO})^{-115/72}}{(\Delta_{\rm i}/r_{\SCO})^{5/2}}
\left[1 - \left(\frac{r_{\rm i}}{r}\right)^{115/72}\left(\frac{\Delta_{\rm i}}{\Delta}\right)^{5/2}\right]\,.
\end{align}
In the second line, the approximation is accurate to within $6\%$ for
$\Delta_{\rm i} \equiv r_{\rm i} - r_{\SCO} \leq 0.3\, r_{\SCO}$.
Note that $T_{\nu}^{\rm neu}$ depends on radius only through $\zeta$;
it increases monotonically and approaches a constant value, which
depends very sensitively on $\Delta_{\rm i}/r_{\SCO}$.  In practice,
one might expect
\begin{equation}\label{e:Deltai}
\Delta_{\rm i}\sim r_{\rm H}~~{\rm and}~~ r_{\rm i}= r_{\SCO}+r_{\rm H}
\end{equation}
for a disk without a { cavity} because the tidal torque model is valid only
outside this region, and within this distance the gas may flow across
the secondary orbit along radial streams or horse shoe orbits.  In the
following we keep $\Delta_{\rm i}/r_{\rm H}$ general.  We incorporate
a factor of $(r_{\rm H}/r_{\SCO})^{-5/2}$ in the prefactor of
Eq.~(\ref{e:Tnuoutunsat4}) and introduce a renormalized $\zeta$, as
\begin{align}
 \zeta_{\rm R} &\equiv \left(\frac{r_{\rm H}}{r_{\SCO}}\right)^{5/2} \zeta\\\nonumber
&\approx
 \left(\frac{2}{3}\right)^{1/6}\frac{2}{5}
\frac{(r_{\rm i}/r_{\SCO})^{-115/72}}{(\Delta_{\rm i}/r_{\rm H})^{5/2}}
\left[1 - \left(\frac{r_{\rm i}}{r}\right)^{115/72}\left(\frac{\Delta_{\rm i}}{\Delta}\right)^{5/2}\right]\,.
\end{align}
Then Eq.~(\ref{e:Tnuoutunsat4}) becomes
\begin{align}
T_{\nu}^{\rm neu} &= 1.7\times 10^{49}{\rm erg}\, \alpha_{-1}^{-2} M_7^{5/2} f_{-2}^{5/2} q_{-3}^{5/2} r_{\SCO 2}^{1/4}
\zeta_{\rm R}^{3}\label{e:Tnuout-unsat}
\\\nonumber
 &\approx T_{\nu \max}^{\rm neu}
\times\left[1 - \left(\frac{r_{\rm i}}{r}\right)^{115/72}\left(\frac{\Delta_{\rm i}}{\Delta}\right)^{5/2}\right]^{3}
\end{align}
where
\begin{align}
T_{\nu \max}^{\rm neu} &\equiv \lim_{r\rightarrow \infty} T_{\nu}^{\rm neu}(r,r_{\SCO}, r_{\rm i})  \\
&\approx 9.1\times 10^{47}{\rm erg}\, \alpha_{-1}^{-2} M_7^{5/2} f_{-2}^{5/2} q_{-3}^{5/2}
 r_{s2}^{1/4} \nonumber\\\quad &\quad\times
\left(\frac{\Delta_{\rm i}}{r_{\rm H}}\right)^{-15/2} \left(1+\frac{\Delta_{\rm i}}{r_{\SCO}}\right)^{-115/24}\,.
\label{e:Tnuout-unsatmax}
\end{align}

The outer edge of this region is where Eq.~(\ref{e:dTnuout}) is first
violated, i.e. where the viscous torque density\footnote{Note that
matching the derivatives at the interface does not contradict
$T_{\nu}^{\rm neu}\gg \dot{M}r^2\Omega$ there.}  becomes comparable to
the accretion term $\partial_r T_{\nu}\sim \dot{M}\partial_r
(r^2\Omega)$.  We substitute $\partial_r T_{\nu}$ from
Eq.~(\ref{e:Tnuoutunsat2}) utilizing (\ref{e:Tnuoutunsat4}) and get
\begin{align}
 1=\frac{\partial_r T_{\nu}^{\rm neu}}{\dot{M} \partial_r (r^2\Omega)} &=
2.8\times 10^{-7} \,\alpha_{-1}^{-2}\dot{m}_{-1}^{-1} M_7^{1/2} f_{-2}^{5/2} q_{-3}^{5} r_{\SCO 2}^{-1/4}
\nonumber\\&\quad\times
\left(\frac{r}{r_{\SCO}}\right)^{-11/24} \left(\frac{\Delta}{r_{\SCO}}\right)^{-7/2} \zeta(r,r_{\SCO}, r_{\rm i})^{2}\,.
\label{e:dTnu_neu}
\end{align}
We label the radial distance of this interface from the secondary as
$\Delta^{\rm neu}_{\rm m}$. While this equation of a single variable
can be easily solved numerically for $\Delta^{\rm neu}_{\rm m}$ for
any fixed $r_{\SCO}$ and $\Delta_{\rm i}$, we may derive an analytical
approximate solution as follows.  Assuming $1\gg\Delta^{\rm neu}_{\rm
m}\gtrsim 3 \Delta_{\rm i}$, $\zeta$ is close to its asymptotic maximum,
which implies
\begin{align}
 \frac{\Delta^{\rm neu}_{\rm m}}{\Delta_{\rm i}} &= 5.0\, \alpha^{-4/7} \dot{m}_{-1}^{-2/7} M_7^{1/7}
f_{-2}^{5/14} q_3^{13/21} r_{\SCO 2}^{-1/14}
\nonumber\\&\quad\times
 \frac{[1+(\Delta_{\rm i}/r_{\SCO})]^{-115/126}}{(\Delta_{\rm i}/r_{\rm H})^{17/7}}\,.
\label{e:Delta_neu}
\end{align}

To obtain the dependence of the physical parameters on radius, we
first derive $D_{\nu}$ and $F$ by substituting
Eq.~(\ref{e:Tnuout-unsat}) into (\ref{e:Dnu}) and
(\ref{e:Foutunsat2}).  Then $\Sigma$ and $T_{\rm c}$ follow from
Eqs.~(\ref{e:Sigma}) and (\ref{e:Tc}).  In the range $r_{\rm i}\leq r
\lesssim r_{\SCO} + \Delta^{\rm neu}_{\rm m}$,
\begin{align}
\Sigma^{\rm neu} &=1.0\times 10^7\frac{\rm g}{\rm cm^2} \alpha_{-1}^{-2} f_{-2}^{3/2}  M_7^{1/2} q_{-3}^{4/3} r_{\SCO 2}^{-3/4}
\left(\frac{r}{r_{\SCO}}\right)^{-23/24}
\nonumber\\&\quad\times
\left(\frac{\Delta}{r_{\SCO}}\right)^{1/2}\zeta_{\rm R}^2\,,\\
T_{\rm c}^{\rm neu} &=6.3\times 10^6{\,\rm K\,} \alpha_{-1}^{-1} f_{-2}  r_{\SCO 2}^{-1} q_{-3}^{7/6}
\left(\frac{r}{r_{\SCO}}\right)^{-25/24}\left(\frac{\Delta}{r_{\SCO}}\right)^{-1/2}\zeta_{\rm R},\\
 F^{\rm neu} &= 6.8\times 10^{12} \frac{\rm erg}{\rm  cm^2\, s} \alpha_{-1}^{-2} f_{-2}^{5/2} q_{-3}^{10/3} M_7^{-1/2}
 r_{\SCO 2}^{-13/4}\nonumber\\ &\quad\times
\left(\frac{r}{r_{\SCO}}\right)^{-77/24}\left(\frac{\Delta}{r_{\SCO}}\right)^{-5/2} \zeta_{\rm R}^{2}\,,
\label{e:Fneu}
\\
v_{r}^{\rm neu} &=173\,\frac{\rm cm}{\rm s^2}\, \alpha_{-1}^{2} \dot{m}_{-1} f_{-2}^{-3/2}  M_7^{-1/2} q_{-3}^{-4/3} r_{\SCO 2}^{-1/4}
\left(\frac{r}{r_{\SCO}}\right)^{-1/24}
\nonumber\\&\quad\times
\left(\frac{\Delta}{r_{\SCO}}\right)^{-1/2}\zeta_{\rm R}^{-2}\,,\label{e:vr_neu}\\
H^{\rm neu} &=
\left\{
 \begin{array}{l}
 0.13 M_{\SMBH}\,\alpha_{-1}^{-2} f_{-2}^{5/2}  M_7^{1/2} r_{\SCO 2}^{-1/4} (r/r_{\SCO})^{-5/24}\\
\quad\times(\Delta/r_{\SCO})^{-5/2}\zeta_{\rm R}^2~~{\rm if~}\beta\approx 0\,, \\
0.23 M_{\SMBH}\,\alpha_{-1}^{-1/2} f_{-2}^{1/2}  r_{\SCO 2} (r/r_{\SCO})^{47/48}
(\Delta/r_{\SCO})^{-1/4}\zeta_{\rm R}^{1/2}\\
\quad{\rm if~}\beta\approx 1\,.
\end{array}
\right.
\end{align}
Here $r=r_{\SCO}+\Delta$, and these equations are formally correct to
first beyond leading order in $\Delta/r_{\SCO}$, but we find them to
be a good approximation typically within $15\%$ even for
$\Delta/r_{\SCO}\gtrsim 1$.  Interestingly, all of these physical
parameters have a local extremum in this zone.  We label the distance
corresponding to the maximum local disk luminosity $4\pi r^2 F(r)$ as
$\Delta_{\rm peak}^{\rm neu}$.  We find that $\Delta_{\rm peak}^{\rm
neu}/\Delta_{\rm i}$ is a slowly decreasing function of $\Delta_{\rm i}$, which
varies between $1.55$ and $1.4$ for $0<\Delta_{\rm i}\lesssim r_{\SCO}$.

To match $T_\nu$ at a radius $r=r_{\SCO}+\Delta_{\rm m}^{\rm neu}$,
the interface between this region and the middle zone, we must set
$T_{\rm bc}\equiv T_{\nu}^{\rm neu}(r_{\rm m}^{\rm neu})$. Based on
Eqs.~(\ref{e:Tnuout-unsat}) we may use
\begin{equation}
T_{\rm bc} \approx T_{\nu \max}^{\rm neu}~~~{\rm if~}\Delta_{\rm m}^{\rm neu}\gg \Delta_{\rm i}\,.
\end{equation}
This equation is typically valid in the overflowing case (see
Eq.~\ref{e:Delta_neu}), as long as $q$ is large enough that $T_{\nu
\max}^{\rm neu} > T_{\nu 0}(r_{\SCO})$ (strongly perturbed solution),
but not too large so that $H < r-r_{\SCO}$. We discuss the solutions
if the latter condition is violated in \S~\ref{s:nearout-sat}
below. Comparing Eqs.~(\ref{e:Tnu0}) and
(\ref{e:Tnuout-unsatmax}) shows that the minimum mass ratio to cause a
significant gas buildup with unsaturated torques:
\begin{equation}\label{e:qneu}
 q^{\rm neu}_{\min} = 9\times 10^{-4} \alpha_{-1}^{4/5} \dot{m}_{-1}^{2/5} M_7^{-1/5} f_{-2}^{-1} r_{\SCO 2}^{1/10}\,.
\end{equation}
For smaller masses, the disk structure is not modified significantly
outside the secondary, and one has to use Eq.~(\ref{e:Tnuin}) for the
torque at the inner boundary in Eq.~(\ref{e:Tnuoutunsat3}) with
$T_{\nu}(r_{\rm i})=T_{\nu}^{\rm ni}(r-r_{\rm H})$. We do not show these
more general but more complicated expressions here.

We note that the results in this section are sensitive to $\Delta_{\rm
i}$ if different from $r_{\rm H}$, which sets the distance at which
the gas can flow in across the secondary's orbit without significant
resistance.  While $\Delta_{\rm i}\sim r_{\rm H} $ is reasonable based
on the horse shoe orbits in the restricted three body problem, we keep
it as a free parameter in the following.

\subsubsection{Outside the secondary -- saturated torque}\label{s:nearout-sat}

Here we again assume that Eqs.~(\ref{e:dTnuout}--\ref{e:Foutdef})
hold, but now examine the case of much higher secondary masses, where
the scale-height is increased so much that $H(r) > r - r_{\SCO}$ and
the tidal perturbation enters the torque cutoff regime.  We make the
simplifying assumption here that $H(r) > r - r_{\SCO}$ holds
throughout the near zone so that the tidal torque in
Eq.~(\ref{e:Lambda}) does not alternate between saturated and
unsaturated.  It is straightforward to obtain more general solutions,
but we find this exclusively saturated OR unsaturated assumption to be
an excellent approximation in most cases.

Due to the large torque barrier and tidal heating, the disk in this
region is typically radiation pressure dominated, and we accordingly
assume $\beta\approx 0$ for the analytical solutions below.

Eqs.~(\ref{e:H}), (\ref{e:Dnu}), and (\ref{e:Sigma}) show that
$H=\frac{\kappa}{\C} \Omega^{-2} F$ and $\Sigma =a_{\Sigma} r^{-8/5}
T_{\nu}^{4/5} F^{-1/5}$ for $b=1$, where $a_{\Sigma}$ is a
constant. Substituting into Eq.~(\ref{e:Foutdef}),
\begin{equation}\label{e:Foutsat2}
F =
\left(\frac{3}{8\pi}\right)^{4/5}\frac{a_{\Sigma}}{4}\frac{\C^4}{\kappa^4}
f q^2 r_{\SCO}^4 r^{2/5}\Omega^{10} (\Omega_{\SCO} - \Omega)
\frac{T_{\nu}^{4/5}}{F^{21/5}}.
\end{equation}
This equation can be solved for $F$ as a function of $r$ and
$T_{\nu}$. Plugging back into Eq.~(\ref{e:dTnuout}) leads to a
separable first order differential equation for $T_{\nu}$
\begin{equation}\label{e:Tnuoutsat2}
\frac{\partial_r T_{\nu}}{T_{\nu}^{2/13}} = a_{1} r_{\SCO}^{9/52}
\left(\frac{\Omega_{\SCO}-\Omega}{\Omega_{\SCO}}\right)^{-21/26}
\left(\frac{\Omega}{\Omega_{\SCO}}\right)^{25/13}
\left(\frac{r}{r_{\SCO}}\right)^{14/13},
\end{equation}
where $a_1$ is a constant independent of $r$ and $r_{\SCO}$. Now
use\footnote{The disk rotates with nearly the local Keplerian angular
  velocity, but slightly slower due to a radial pressure gradient:
  $(\Omega_{\rm K}-\Omega)/\Omega_{\rm K}\propto H^2/r^2$ see Eq.~(78)
  in \citet{2011PhRvD..84b4032K}.}  $\Omega/\Omega_{\SCO}\approx
(r/r_{\SCO})^{-3/2}$ and integrate both sides assuming
$T_{\nu}(r_{\rm i})\approx 0$ for some $r_{\rm i} \gtrsim r_{\SCO}$. Here $r_{\rm i}$ is
the radius where the torque model breaks down, for which we adopt the
Hill radius around the secondary $r_{\rm i} = [1+(q/3)^{1/3}] r_{\SCO}$ if a
gap does not form. Thus,
\begin{equation}\label{e:Tnuout3}
T_{\nu}^{11/13}(r) - T_{\nu}^{11/13}(r_{\rm i}) =\frac{11}{13} a_{1}
r_{\SCO}^{61/52} \psi(r,r_{\SCO},r_{\rm i}),
\end{equation}
where we introduced a dimensionless function
\begin{align}\label{e:psi}
\psi &= \frac{2}{3}\Beta\hspace{-2pt}\left(\frac{\Omega_{\SCO}-\Omega}{\Omega_{\SCO}};\frac{5}{26},\frac{7}{13}\right)
-\frac{2}{3}\Beta\hspace{-2pt}\left(\frac{\Omega_{\SCO}-\Omega_{0}}{\Omega_{\SCO}};\frac{5}{26},\frac{7}{13}\right)\\
&\approx 3.7 \left[ \left(\frac{r-r_{\SCO}}{r}\right)^{5/26} - \left(\frac{r_{\rm i}-r_{\SCO}}{r_{\rm i}}\right)^{5/26}\right]\label{e:psi1}\\
&\approx 0.4 \ln \frac{ r-r_{\SCO}}{ r-r_{\rm i}}.\label{e:psi2}
\end{align}
Here $\Beta(x;a,b)=\int_0^x t^{a-1} (1-t)^{b-1} \D t$ is the
incomplete beta function, and the last two lines are simple
approximations, typically accurate to within $15\%$.  We can now use
Eqs.~(\ref{e:Dnu}) to get $D_{\nu}$. For $T_{\nu}(r_{\rm i})\approx 0$,
after substituting the value of $a_1$, Eq.~(\ref{e:Tnuout3}) yields
\begin{align}\label{e:Tnuout4}
T_{\nu}^{\rm nes} &= 1.1\times 10^{49}{\rm erg}\, \alpha_{-1}^{-2/11} f_{-2}^{5/22} q_{-3}^{5/11} M_7^{45/22}
r_{\SCO 2}^{61/44} \psi^{13/11}\\
D_{\nu}^{\rm nes} &= 1.2\times 10^{15}\frac{{\rm erg}}{\rm cm^2 s}\, \alpha_{-1}^{-2/11} f_{-2}^{5/22} q_{-3}^{5/11} M_7^{-21/22}
\nonumber\\
&\quad\times r_{\SCO 2}^{-93/44} \left(\frac{r}{r_{\SCO}}\right)^{-7/2}\psi^{13/11}\,.\label{e:Dnuout}
\end{align}
Here the superscript $^{nes}$ refers to the case of saturated torque
in the external near zone. Note that $T_{\nu}^{\rm nes}$ depends on
radius only through $\psi$. Close to the inner boundary of this region
$r_{\rm i}$, it grows quickly with $\delta r/r$ and saturates to a
constant at $\delta r/r\sim 1$.  This can be understood, since this
solution neglects angular momentum flow, the viscous torque is equal
to the integrated tidal torque density, and the latter has a cutoff at
$\delta r/r\sim 1$.  We can verify that $T_{\nu}^{\rm nes}\gg T_{\nu
0}$ is indeed satisfied in this region (c.f. Eq.~\ref{e:Tnu0}) and so
the first assumption, Eq.~(\ref{e:dTnuout}), and $T_{\nu}(r_{\rm
i})\approx 0$ are well justified.

Can we use the asymptotic maximum of $T_{\nu}^{\rm nes}$ as an
estimate of the torque at the outer boundary of this region, to
estimate $T_{\rm bc}^{\rm mo}$ in the middle zone of an overflowing
disk?  In many cases no, because the disk transitions to the middle
zone much closer $\delta r \ll r_{\SCO}$, implying that the torque at
the outer boundary of this region can be much less than its asymptotic
maximum.  The outer boundary of this region is where
$H=r-r_{\SCO}$. To figure out exactly where this happens, we proceed
to determine the disk structure in this region.

Now Eqs.~(\ref{e:Dnuout}) and (\ref{e:Foutsat2}) give $D_{\nu}$ and
$F$; all other disk parameters then follow from Eq.~(\ref{e:mdot}),
(\ref{e:H}), and (\ref{e:Sigma}--\ref{e:Tc}). The result within $r_{\rm i}
\leq r \lesssim r_{\SCO} + H^{\rm nes}$ is
\begin{align}
\Sigma^{\rm nes} &= 2.6\times 10^6 \frac{\rm g}{{\rm cm}^2} \alpha_{-1}^{-10/11} M_7^{5/22}  f_{-2}^{3/22} q_{-3}^{3/11} r_{\SCO 2}^{-3/44}\nonumber\\
&\quad \times \left(\frac{\Omega_{\SCO}-\Omega}{\Omega_{\SCO}}\right)^{-1/26} \left(\frac{r}{r_{\SCO}}\right)^{-27/26} \psi^{10/11}\,,\\
T_c^{\rm nes} &= 1.6\times 10^6{\,\rm K}\,\alpha_{-1}^{-3/11} M_7^{-2/11}  f_{-2}^{1/11} q_{-3}^{2/11} r_{\SCO 2}^{-6/11}\nonumber\\
&\quad \times \left(\frac{\Omega_{\SCO}-\Omega}{\Omega_{\SCO}}\right)^{1/26}
\left(\frac{r}{r_{\SCO}}\right)^{-25/26} \psi^{3/11}\,,\\
F^{\rm nes} &= 9.8\times 10^{14}\frac{\rm erg}{{\rm cm}^2{\rm s}}\,\alpha_{-1}^{-2/11}  M_7^{-21/22} f_{-2}^{5/22} q_{-3}^{5/11}  r_{\SCO 2}^{-93/44}\nonumber\\
&\quad \times \left(\frac{\Omega_{\SCO}-\Omega}{\Omega_{\SCO}}\right)^{5/26}
\left(\frac{r}{r_{\SCO}}\right)^{-73/26} \psi^{2/11}\,,\label{e:Fout}\\
v_r^{\rm nes} &= 660\frac{\rm cm}{\rm s}\,\alpha_{-1}^{10/11} \dot{m}_{-1} M_7^{-5/22} f_{-2}^{-3/22}   q_{-3}^{-3/11} r_{\SCO 2}^{-41/44}\nonumber\\
&\quad \times \left(\frac{\Omega_{\SCO}-\Omega}{\Omega_{\SCO}}\right)^{1/26}
\left(\frac{r}{r_{\SCO}}\right)^{1/26} \psi^{-10/11}\,,\label{e:vrout}\\
H^{\rm nes} &= 19\,M_{\SMBH}\, \alpha_{-1}^{-2/11} M_7^{1/22}  f_{-2}^{5/22} q_{-3}^{5/11} r_{\SCO 2}^{39/44} \nonumber\\
&\quad \times \left(\frac{\Omega_{\SCO}-\Omega}{\Omega_{\SCO}}\right)^{5/26}
\left(\frac{r}{r_{\SCO}}\right)^{5/26} \psi^{2/11}\,.
\label{e:Hout}
\end{align}
We can now confirm the consistency of the second assumption,
Eq.~(\ref{e:Foutdef}), using Eqs.~(\ref{e:Dnuout}) and
(\ref{e:Fout}). Indeed, $D_{\nu}^{\rm nes}\lesssim F^{\rm nes}$ near
the secondary since $D_{\nu}^{\rm nes}$ scales with a higher power of
$\psi$.  $\Sigma$ and $T_{c}$ have a maximum, $v_r$ decreases and
becomes practically constant, while $H$ slowly increases in this
regime.

The outer boundary of this region, $r^{\rm nes}_{\rm m}$, is
where\footnote{Note that in this section we use $\delta r\equiv
r-r_{\SCO}$ instead of $\Delta$ since in this region
$\Delta=\max(\delta r, H)=H$ (see Eq.~\ref{e:Delta}).}
\begin{equation}\label{e:Hout1}
 H^{\rm nes}(r^{\rm nes}_{\rm m}) = \delta r^{\rm nes}_{\rm m} \equiv r^{\rm nes}_{\rm m}-r_{\SCO}\,.
\end{equation}
We use $r^{\rm nes}_{\rm m}$ as an approximation to the transition
radius to the middle zone.  After substituting Eq.~(\ref{e:Hout}),
Eq.~(\ref{e:Hout1}) is a nonlinear algebraic equation for $r^{\rm
nes}_{\rm m}$.  While it is easy to solve it numerically for any
choice of parameters, it is still useful to derive approximate
analytical solutions.  We find the following method yields results
that are accurate within $20\%$ for a wide range of parameters.  Use
Eq.~(\ref{e:psi1}), expand $H^{\rm nes}$ to second order in $\delta
r^{\rm nes}_{\rm m}$, use $(1+a x) \approx (1+x)^a $ for small $x$ and
$1-x^a \approx -a\ln x$ for small $a$. This gives an approximate
relation
\begin{align}
 \frac{\delta r^{\rm nes}_{\rm m}}{r_{\SCO}} &\approx 0.17\,\alpha_{-1}^{-4/17}M_7^{1/17}  f_{-2}^{5/17} q_{-3}^{10/17} r_{\SCO 2}^{-5/34}
\nonumber\\&\times \left( \frac{5}{26} \ln \frac{\delta  r^{\rm nes}_{\rm m}}{\delta r_{\rm i}}   \right)^{4/17}\,,
\label{e:dr1}
\end{align}
where $\delta r_{\rm i} \equiv r_{\rm i} - r_{\SCO}$.
Eq.~(\ref{e:dr1}) can be solved analytically using the Lambert
W-function \citep{springerlink:10.1007/BF02124750}\footnote{ The
Lambert W-function is defined to be the inverse of the function
$f(\W)=\W \exp(\W)$, where we need the real branch with the larger
absolute value, $\W_{-1}$, defined on $f>-e^{-1}=-0.368$. The
approximation in Eq.~(\ref{e:W}) is correct to within $20\%$ for all
$0<a<1/e$.}
\begin{equation}\label{e:W}
 \W_{-1}(-a) \approx \ln(a) - \ln(-\ln(a))
\end{equation}
as
\begin{equation}\label{e:dr1b}
 \delta r^{\rm nes}_{\rm m} = \delta r_{\rm i} \left[\frac{\W_{-1}(-a)}{-a}\right]^{4/17} = \delta r_{\rm i} \exp\left[-\frac{4}{17} \W_{-1}(-a)\right]\,,
\end{equation}
where the two forms are equivalent, and we have introduced
\begin{equation}\label{e:adef}
 a = 0.465\, \alpha_{-1} M_7^{-1/4}  f_{-2}^{-5/4} q_{-3}^{-13/12} r_{\SCO 2}^{5/8} \left(\frac{\delta r_{\rm i}}{r_{\rm H}}\right)^{17/4}\,.
\end{equation}
Finally, we substitute in Eq.~(\ref{e:Tnuout4}) and use Eq.~(\ref{e:psi2}),
\begin{align}
T_{\nu}^{\rm nes}(r^{\rm nes}_{\rm m}) &= 6.9\times 10^{47}{\rm erg}\, \alpha_{-1}^{-2/11} M_7^{45/22} f_{-2}^{5/22} q_{-3}^{5/11}
r_{\SCO 2}^{61/44}\nonumber\\
&\quad\times \left[-\W_{-1}(-a)\right]^{13/11}\,.  \label{e:Tnuout5}
\end{align}
Note that $\W_{-1}(-a)$ depends logarithmically weakly on the disk
parameters \citep{springerlink:10.1007/BF02124750}, in practice $1\leq
|\W_{-1}(-a)|\lesssim 10$ for $10^{-4}\lesssim a \leq 1/e=0.368$.
Here $a\leq 1/e$ is required for this solution to exist, implying that
$q$ and $r_{\SCO}$ have to be sufficiently large and small,
respectively.  In the opposite case, the torque is unsaturated
(\S~\ref{s:nearout-un}).

\subsection{Transition between near and middle zones}\label{s:nearout-middle}
\subsubsection{The case with overflow}

The value of $T_{\nu}$ at the outer edge of the near zone, is to be
matched with that in the middle zone, $T_{\rm bc}$.  If the tidal
torque is unsaturated in the near zone, we approximate $T_{\rm bc}$
with the asymptotic maximum value, $T_{\rm bc}=T_{\nu \max}^{\rm
neu}$. Otherwise, if it is saturated, then we set $T_{\rm
bc}=T_{\nu}^{\rm nes}(r^{\rm nes}_{\rm m})$. Matching the middle and
near zones at $r^{\rm nes}_{\rm m}$ assumes that $T_{\nu}$ does not
grow substantially in the transition region between the saturated near
zone and the middle zone, i.e.  outside of $r^{\rm nes}_{\rm m}$ but
within a radius where the tidal effects are still non-negligible.  We
find this approximation to be better than $10\%$.  Thus, to match the
torque at the outer boundary of the near zone and the inner boundary
of the overflowing middle zone, we combine the saturated and
unsaturated cases as
\begin{equation}\label{e:c0mo}
 T_{\rm bc}^{\rm mo} = \left\{
\begin{array}{ll}
\min\{T_{\nu}^{\rm nes}(r^{\rm nes}_{\rm m}), T_{\nu \max}^{\rm neu}\}       &{\rm if~}a\leq 0.368\,, \\
T_{\nu \max}^{\rm neu}    &{\rm if~} a\geq 0.368\,,
\end{array}
\right.\,\vspace{0pt}
\end{equation}
where $T_{\nu}^{\rm nes}(r^{\rm nes}_{\rm m})$ and $T_{\nu \max}^{\rm
neu}$ are given by Eqs.~(\ref{e:Tnuout-unsatmax}) and
(\ref{e:Tnuout5}).  If this satisfies $T_{\rm bc}^{\rm mo} < T_{\rm
bc}^{\rm mg}$ for $\delta r_{\rm i}=r_{\rm H}$ (see Eq.~\ref{e:c0}),
then the satellite migration velocity is less than the gas bulk local
inflow velocity, and the overflowing steady-state solution is
self-consistent. In the opposite case the disk forms a gap with
$\delta r_{\rm i}>r_{\rm H}$.

The dimensionless { angular momentum flux} in the middle zone
(Eq.~\ref{e:k}) is
\begin{align}\label{e:kmou}
k^{\rm mou}_{\SCO} &= 1.3\, \alpha_{-1}^{-2} \dot{m}_{-1}^{-1} M_7^{1/2} f_{-2}^{5/2} q_{-3}^{5/2} r_{\SCO 2}^{-1/4} \left[1+\left(\frac{q}{3}\right)^{1/3}\right]^{-15/2}\,,\\
\label{e:kmos}
k^{\rm mos}_{\SCO} &= 0.97\, \alpha_{-1}^{-2/11} \dot{m}_{-1}^{-1} M_7^{1/22} f_{-2}^{5/22} q_{-3}^{5/11} r_{\SCO 2}^{39/44} |\W(a)|^{13/11}\,.
\end{align}
Gap overflow causes the torque level to decrease in the middle zone,
which suppresses $k$. We discuss gap closing in
\S~\ref{s:gap} below.

The disk flux in the near zone is larger, due to tidal heating, than
in the middle zone.  To show this, we next compare the luminosity of
the near zone to the middle zone explicitly.  In the case of
unsaturated torques in the near zone, we find that the local disk
luminosity, $L^{\rm neu}(r)\equiv4\pi r^2 F^{\rm neu}$ (see discussion
following Eq.~\ref{e:Fneu}), peaks sharply near $r^{\rm neu}_{\rm
peak} = r_{\SCO} + 1.5\, r_{\rm H}$. We find that the integrated flux
from within a ring of width $0.5\,r_{\rm H}$ is approximately
\begin{equation}
k^{\rm neu}_{\SCO} = 1.0 \frac{r_{\rm H}}{r_{\SCO}} k^{\rm mou}_{\SCO}\,.
\end{equation}
In the unsaturated case, $r_{\rm H}/r_{\SCO}= (q/3)^{1/3}\ll 1$, and
so the net luminosity of the middle zone exceeds that of the near
zone.  For saturated torques, the maximum brightness corresponds to
the outer boundary of the near zone, $r^{\rm nes}_{\rm m}$. We assume
an effective radial width $\delta r^{\rm nes}_{\rm m}$ given by
Eq.~(\ref{e:dr1}). From Eqs.~(\ref{e:H}) and (\ref{e:Hout1}), assuming
a radiation pressure dominated near zone, the { dimensionless angular momentum flux} of
the near zone relative to the unperturbed disk is
\begin{align}\label{e:knes}
k^{\rm nes}_{\SCO}
&\equiv
\frac{4\pi r^{\rm nes}_{\rm m} \delta r^{\rm nes}_{\rm m} F^{\rm nes}(r^{\rm nes}_{\rm m})}{4\pi (r^{\rm nes}_{\rm m})^2 F_{0}(r_1)}
= \frac{\delta r^{\rm nes}_{\rm m}}{r^{\rm nes}_{\rm m}} \frac{H^{\rm nes}(r^{\rm nes}_{\rm m})}{H_{0}}
= \frac{(\delta r^{\rm nes}_{\rm m})^2}{r^{\rm nes}_{\rm m} H_{0}}\\\nonumber
&= 0.45\, \alpha_{-1}^{-8/17} \dot{m}^{-1} f_{-2}^{10/17} M_7^{2/17} q_{-3}^{20/17} r_{\SCO 2}^{12/17}[-\W(a)]^{8/17}\,.
\end{align}
Comparing Eqs.~(\ref{e:kmos}) and (\ref{e:knes}), and recalling that
typically $1\leq |\W(-a)|\lesssim 10$, we conclude that the luminosity
of the saturated near zone can exceed that of the middle zone by a
factor between $\sim 3-10$ for $q\sim 0.1$.

Given the { dimensionless angular momentum flux} in the middle zone, the disk parameters
and the migration rate are given by
Eqs.~(\ref{e:Sigmam}--\ref{e:vSCOrm}).  We substitute
Eqs.~(\ref{e:kmou}--\ref{e:kmos}) to obtain an explicit formula for
the disk parameters.  The migration rate in the overflowing disk with
saturated and unsaturated torques is, respectively,
\begin{align}\label{e:vSCO_ou}
 v_{\SCO r}^{\rm ou} &= 30 \frac{\rm cm}{\rm s} \alpha_{-1}^{-2} f_{-2}^{5/2} q_{-3}^{3/2} M_7^{3/2} r_{\SCO 2}^{3/4}
\left[1+\left(\frac{q}{3}\right)^{1/3}\right]^{-115/24}\,,\\
 v_{\SCO r}^{\rm os} &= 23 \frac{\rm cm}{\rm s} \alpha_{-1}^{-2/11} f_{-2}^{5/22} q_{-3}^{-6/11} M_7^{23/22} r_{\SCO 2}^{83/44}
[-\W(a)]^{13/11}\,.
\label{e:vSCO_os}
\end{align}

\subsubsection{Disk with a cavity}\label{s:nearout-gap}

Let us next turn to the case with a gap. We determine the radial
distance to the outer edge of the gap, $\lambda r_{\SCO}$, here by
requiring that the migration velocity matches the rescaled gas inflow
velocity in the middle zone as stated in the boundary condition,
Eq.~(\ref{e:vrdef}). With $\lambda$ in hand,
Eq.~(\ref{e:Sigma2b}--\ref{e:vSCOr_mg}) determine the disk parameters
in the middle zone.  The solution is different when the tidal torque
is unsaturated near the inner edge and when it is saturated, which we
discuss in turn below.

First, assume that the torque is unsaturated all the way outside of
the gap ($H \leq r-r_{\SCO}$).  We use Eq.~(\ref{e:dTnu_neu}) to
obtain $\zeta(\lambda r_{\SCO},r_{\SCO},r_{\rm i})$ at the interface
between the near and the middle zone. We substitute in
(\ref{e:vr_neu}) to obtain the gas velocity at the interface rescaled
by $1/\lambda$:
\begin{equation}\label{e:vrneu1}
 \frac{v_{r}^{\rm neu}(\lambda r_{\SCO})}{\lambda} = 30\frac{\rm
cm}{\rm s}\, f_{-2} q_{-3}^{2} r_{\SCO 3}^{-1/2}  \lambda^{-3/2}
(\lambda-1)^{-4}.
\end{equation}
To obtain the migration velocity, we identify $T_{\rm bc}=T_{\nu}^{\rm
neu}(\lambda r_{\SCO})$ in Eq.~(\ref{e:vSCOrm}), and substitute
Eq.~(\ref{e:Tnuoutunsat4}), and eliminate $\zeta$ using
Eq.~(\ref{e:dTnu_neu}). This gives
\begin{equation}\label{e:vrneu2}
 v_{\SCO r} = 8022\frac{\rm cm}{\rm s}\, \alpha_{-1}
\dot{m}_{-1}^{3/2} M_{7}^{3/4} f_{-2}^{-5/4} q_{-3}^{-7/2} r_{\SCO
2}^{9/8} \lambda^{11/16} (\lambda-1)^{21/4}.
\end{equation}
The boundary condition, Eq.~(\ref{e:vrdef}), states that
Eqs.~(\ref{e:vrneu1}) and (\ref{e:vrneu2}) must be equal.  This
provides a nonlinear equation for $\lambda$. We solve this equation
perturbatively. To first beyond leading order,
\begin{equation}
 \lambda_{\rm u} = 1+ \delta_{\rm u}(1+\delta_{\rm u})^{-35/148},
\end{equation}
where
\begin{align}
 \delta_{\rm u} = 0.55\, \alpha_{-1}^{-4/37} \dot{m}^{-6/37}
 f_{-2}^{9/37} q_{-3}^{22/37} M_{7}^{-3/37} r_{\SCO 2}^{-13/74}.
\end{align}
The `$\rm u$' subscript is introduced to distinguish the case with
unsaturated torques.

Next, consider the case of the torque cutoff.  The formulas in
\S~\ref{s:nearout-sat} are not limited to the overflowing case, as
long as the torque is saturated ($\delta r\equiv r-r_{\SCO} \leq H$).
If a gap opens then $\delta r_{\rm i}$ marks the distance to the edge
of the disk in Eq.~(\ref{e:Tnuout3}--\ref{e:psi}), for which $\delta
r_{\rm i} > r_{\rm H}$.  Here, $\delta r_{\rm i}$ can be eliminated
using the boundary condition Eq.~(\ref{e:vrdef}) as follows. We
identify $\lambda =r^{\rm nes}_{\rm m}/r_{\SCO}$ in
Eq.~(\ref{e:vrdef}) where $r^{\rm nes}_{\rm m}$ marks the edge of the
near zone according to Eq.~(\ref{e:Hout1}), so that
$\lambda=1+\delta_1$ where $\delta_1 = \delta r^{\rm nes}_{\rm
m}/r_{\SCO}$.  Combine Eqs.~(\ref{e:vrout}--\ref{e:Hout1}) to
eliminate $\psi(r_1)$ from the bulk gas velocity at $r_1$
\begin{equation}\label{e:vrout2}
 \frac{v_r^{\rm nes}(r^{\rm nes}_{\rm m})}{\lambda} = 0.15\frac{\rm
cm}{\rm s} \dot{m}_{-1} f_{-2} q_{-3}^2 r_{\SCO 2}^{-3/2} \frac{[1 -
(1+\delta_1)^{-3/2}]}{ \delta_1^{5}}.
\end{equation}
Similarly, assuming $T_{\rm bc}=T_{\nu}^{\rm nes}(r^{\rm nes}_{\rm m})$ in
Eq.~(\ref{e:vSCOrm}), eliminate $\psi(r^{\rm nes}_{\rm m})$ from
Eqs.~(\ref{e:Tnuout4}) and (\ref{e:Hout1}), we get
\begin{equation}\label{e:vSCOrout2}
 v_{\SCO r} = 2.0\times 10^7\frac{\rm cm}{\rm s} \alpha_{-1}
f_{-2}^{-5/4} q_{-3}^{-7/2} r_{\SCO 2}^{21/8} \frac{
(1+\delta_1)^{-5/4} \delta_1^{13/2}}{{[1 -
(1+\delta_1)^{-3/2}]^{5/4}}}.
\end{equation}
The boundary condition, Eq.~(\ref{e:vrdef}), states that
Eqs.~(\ref{e:vrout2}) and (\ref{e:vSCOrout2}) are equal.  After
rearranging, we get
\begin{align}\label{e:delta1}
 \delta_1 &= 0.13\, \alpha_{-1}^{-4/37} \dot{m}_{-1}^{4/37}
f_{-2}^{9/37} q_{-3}^{22/37} M_7^{-3/37} r_{\SCO 2}^{-33/74}
\nonumber\\
&\quad\times \left[ \frac{3 + 3\delta_1 + \delta_1^2
}{1+(1+\delta_1)^{3/2}}\right]^{9/37} (1+\delta_1)^{-17/74}\,.
\end{align}
The last two terms can be omitted within $10\%$ accuracy for
$0<\delta<3$.  This gives the characteristic gap scale in the torque
cutoff zone
\begin{equation}\label{e:lambdadef0b}
 \lambda_{\rm s} = 1+0.13\, \alpha_{-1}^{-4/37} \dot{m}_{-1}^{4/37}
 M_7^{-3/37} f_{-2}^{9/37} q_{-3}^{22/37} r_{\SCO 2}^{-33/74},
\end{equation}
provided that $\lambda_{\rm s}-1> r_{\rm H}/r_{\SCO}$
(i.e. $r-r_{\SCO} > r_{\SCO}+r_{\rm H}$); otherwise no gap is
possible.

Thus, the { dimensionless angular momentum flux} follows after substituting into
Eq.~(\ref{e:kmg}). In the unsaturated and saturated cases,
\begin{align}\label{e:kmgu}
 k^{\rm mgu}_{\SCO} &= 23\, \alpha_{-1}^{1/2} \dot{m}_{0.1}^{-3/8}
 M_7^{-3/4} q_{-3}^{5/8} \lambda_{\rm u}^{-11/16} r_{\SCO
 2}^{-7/8}\,,\\
k^{\rm mgs}_{\SCO} &= 23\, \alpha_{-1}^{1/2} \dot{m}_{0.1}^{-3/8}
 M_7^{-3/4} q_{-3}^{5/8} \lambda_{\rm s}^{-11/16} r_{\SCO 2}^{-7/8}\,.
\label{e:kmgs}
\end{align}
The migration speed of the secondary in case of a gap with unsaturated
and saturated tidal torques, respectively, is
\begin{align}\label{e:vSCO_gu}
v_{\SCO r}^{\rm gu} &=
  -550\frac{\rm cm}{\rm s} \alpha_{-1}^{1/2} \dot{m}_{-1}^{5/8} M_7^{1/4} q_{-3}^{-3/8} \lambda_{\rm u}^{-11/16}
r_{\SCO 2}^{1/8}\,,\\
v_{\SCO r}^{\rm gs} &=
  -550\frac{\rm cm}{\rm s} \alpha_{-1}^{1/2} \dot{m}_{-1}^{5/8} M_7^{1/4} q_{-3}^{-3/8} \lambda_{\rm s}^{-11/16}
r_{\SCO 2}^{1/8}\,.
\label{e:vSCO_gs}
\end{align}

These estimates depend on the somewhat arbitrary definition of
$\lambda$ that we have adopted in the two cases.  In the unsaturated
case, we have identified it with the outer edge of the transition
region between the near and middle zones, where $\partial_r T_{\nu} =
\dot{M} \partial_r (r^2\Omega)$, while in the saturated case, we
considered it to be the inner edge of the transition region, where
$H=r-r_{\SCO}$.  While these conventions could be modified, they do
not affect the overflowing solution. They do, however, influence the
secondary orbital radius where the gap closes, which we discuss next.

\subsection{Gap opening and closing}\label{s:gap}

The previous sections define the disk uniquely, which constitute the
solution to the basic equations of \S~\ref{s:interaction}.  By looking
at the solution, we can identify cases where a { cavity is kept empty} in
steady-state or when the disk overflows.

The { cavity refills} if the inner edge of the disk outside the secondary,
$r_{\rm i}$, falls within the Hill radius of the secondary.  The
viscous torque in the near zone, either $T_{\nu}^{\rm nes}$ or
$T_{\nu}^{\rm neu}$, is a monotonically decreasing function of $r_{\rm
i}$.  Thus, if { the disk is truncated}, then the viscous torque at any radius in
the near zone is decreased relative to its value for an overflowing
disk, $r_{\rm i}=r_{\SCO}+r_{\rm H}$ (Eq.~\ref{e:Deltai}).  This shows
that the state of the disk is uniquely determined by the smallest
torque barrier\footnote{recall that $T_{\rm bc}=T_{\nu}$ at the
interface between the near and middle zone}, $T_{\rm bc}$, or
equivalently, the smallest { dimensionless angular momentum flux}:
\begin{equation}
 k_{\SCO}= \min\{k^{\rm mou}_{\SCO},k^{\rm mos}_{\SCO},k^{\rm mgu}_{\SCO},k^{\rm mgs}_{\SCO}\}\,.
\end{equation}
given by Eqs.~(\ref{e:kmou}--\ref{e:kmos}) and
(\ref{e:kmgu}--\ref{e:kmgs}).  We therefore must distinguish four
different possible cases of migration and disk behavior. First,
$k_{\SCO}=k^{\rm mgu}_{\SCO}$ corresponds to the standard case with a
wide gap, with the secondary exhibiting Type-II migration.  If
$k_{\SCO}=k^{\rm mgs}_{\SCO}$, then the gap edge is located within a
scale-height in the near zone so that the torque cutoff limits the
tidal torques, but they can nevertheless support a gap against
viscosity as the secondary migrates inward.  However, if
$k_{\SCO}=k^{\rm mos}_{\SCO}$, then the saturated tidal torque becomes
smaller than the viscous torque all the way to the Hill radius, and
the { cavity refills}.  Finally, if $k_{\SCO}=k^{\rm mou}_{\SCO}$, then the
disk reaches the Hill radius and overflows already while the torques
in the near zone are still unsaturated.

The migration rate of the secondary is proportional to $k_{\SCO}$,
(Eq.~\ref{e:vSCOrm}), i.e.
\begin{equation}
|v_{\SCO r}|=2k_{\SCO} \dot{M} r_{\SCO}/m_{\SCO} =
\min\{v_{\SCO r}^{\rm ou}, v_{\SCO r}^{\rm os}, v_{\SCO r}^{\rm gu},v_{\SCO r}^{\rm gs}\}\,.
\end{equation}
given by Eqs.~(\ref{e:vSCO_ou}--\ref{e:vSCO_os}) and
(\ref{e:vSCO_gu}--\ref{e:vSCO_gs}).  Note that $v_{\SCO r}^{\rm gu}$
and $v_{\SCO r}^{\rm gs}$ are given by practically the same formula,
up to an order-of-unity factor of $\lambda$. This corresponds to the
case of secondary-dominated Type-II migration
\citep{1995MNRAS.277..758S}.  The migration rate in the overflowing
case is $v_{\SCO r}^{\rm ou}$ or $v_{\SCO r}^{\rm os}$ for unsaturated
or saturated torques. Here the disk is still strongly perturbed, but
gas inflow across the orbit limits the efficiency of migration.  The
disk structure in this new regime is intermediate between a disk with
{ an empty} gap (normally associated with Type-II migration) and a weakly
perturbed disk (Type-I migration).  Although the migration speed in
this regime is slower than either in standard Type-II or Type-I
migration, we refer to this regime as ``Type-1.5''.

The gap can also close due to three dimensional overflow for large
secondary masses if the tidal heating is substantial to make the disk
puff up.  Eq.~(\ref{e:H2b}) shows that this happens ($H\gtrsim r$) in
a radiation pressure dominated disk if
\begin{equation}
 r_{\SCO} \lesssim r_{\SCO}^{\rm thick} = 57M_{\SMBH}\, \alpha_{-1}^{4/15} \dot{m}_{-1}^{1/3} M_7^{-2/5} \lambda^{-7/6}
 q_{-3}^{1/3}
\end{equation}
or equivalently if
\begin{equation}
 q \gtrsim q_{\rm thick} =5.4\times 10^{-3}\, \alpha_{-1}^{-4/5} \dot{m}_{-1}^{-1} M_7^{6/5} \lambda^{7/2} r_{\SCO 2}^{3}\,.
\end{equation}
In this case, we do not derive the geometrically thick overflowing
disk or the migration rate.

We discuss the gap opening and closing conditions in more detail in
Paper~II, where we contrast them explicitly with the standard
expressions widely used in the literature, and also compare Type-1.5
migration to the previously known Type-I and II cases.

\section{Results and discussion}\label{s:analytical}

We have derived analytical solutions to the disk model in different
radial regions, where either the tidal, the viscous torques, or the
angular momentum flux is negligible relative to the other two terms in
Eq.~(\ref{e:Tnu'}).  In particular, we have identified the \emph{far
zones}, either well inside or outside the secondary's orbit, where the
effects of the secondary are negligible, the exterior \emph{middle
zone}, where the disk structure is greatly modified but where the
tidal torque and heating are locally negligible, and the \emph{near
zones} just inside or outside the secondary, where the tidal effects
dominate\footnote{We have further restricted the near zones to lie
outside the Hill radius of the secondary.}.  We distinguish two cases
in the middle zone, depending on whether the disk has a gap (i.e.  the
disk is truncated well outside the Hill radius) or if the disk is
overflowing across the secondary's orbit.  We also distinguish two
cases in the near zone outside the secondary's orbit, depending on
whether the tidal torque from the binary is saturated (i.e. whether
the location of the cavity edge falls within a scale height $H$ from
the secondary).  Furthermore, we have investigated asymptotic results
for gas and radiation pressure dominated cases.

Distinguishing all of the above cases allowed us to adopt separate
approximations, each valid in the corresponding regime, and to derive
analytical results to the perturbed accretion disk interacting with
the secondary. We further used this to estimate the migration speed
for the secondary.

In this section, we collect all of the resulting analytical solutions
for the most important disk parameters in the various zones, and
present them in a form suitable for easy use. We refer the reader to
Paper~II for physical interpretations, and discussions on possible
implications of our results for real binary systems.

\begin{table*}
\begin{tabular}{lcrcccccccc}
\hline\hline
			&						&	cgs value		& $[\alpha_{-1}]$& $[\dot{m}_{-1}]$& $[M_7]$	& $[r_{2}]$ & $[f_{-2}]$ & $[q_{-3}]$	& $[r_{\SCO 2}]$ & $\Phi$  \\
\hline
Far zone 			&  $\Sigma_{0}$		& 	$4.7(+5)$		& $-4/5$	& $3/5$	& $1/5$	& $-3/5$		& $0$     	& $0$ 	 & $0$ 		& $\varphi^{3/5}$\\
Middle with gap	& $\Sigma^{\rm mg}$	&	$3.6(+6)$		& $-1/2$	& $3/8$	& $-1/4$	& $-9/10$ 	& $0$     	& $3/8$	 & $-9/40$	 & $\lambda^{-33/80}$\\
Middle w/o gap sat. & $\Sigma^{\rm mos}$	&	$4.6(+5)$		& $-10/11$ & $0$	& $5/22$	& $-9/10$ 	& $3/22$  & $3/11$  & $183/220$	& $|\W|^{13/11}$\\
Middle w/o gap uns.& $\Sigma^{\rm mou}$	&	$5.5(+5)$		& $-2$     & $0$	& $1/2$	& $-9/10$ 	& $3/2$   	 & $3/2$   & $3/20$		& $(\delta r_{\rm i}/r_{\rm H})^{-9/2}$\\
Near ext. sat.		& $\Sigma^{\rm nes}$	&	$2.6(+6)$		& $-10/11$ & $0$	& $5/22$	& $-27/26$  	& $3/22$  & $3/11$	 & $-3/44$	& $\delta\Omega^{-1/26}\psi^{10/11}$ \\
Near ext. uns.		& $\Sigma^{\rm neu}$	&	$1.0(+7)$		& $-2$	   & $0$	& $1/2$	& $-23/24$ 	& $3/2$  	 & $4/3$	 & $5/24$		& $(\delta r/r_{\SCO})^{1/2}\zeta_{\rm R}^2$ \\
Near int.			& $\Sigma^{\rm ni}$	&	$5.7(+7)$		& $0$	   & $1$	& $0$	& $-1/2$  	& $-1$	 & $-2$	 & $0$	 	& $|\delta r/r|^4$ \\
\hline%
Far zone			&  $T_{c0}$			& 	$5.4(+5)$		& $-1/5$ & $2/5$	& $-1/5$	& $-9/10$  	& $0$ 	& $0$ 	& $0$		& $\varphi^{2/5}$\\
Middle with gap	& $T_{c}^{\rm mg}$		&	$1.9(+6)$		& $0$	 & $1/4$	& $-1/2$	& $-11/10$ 	& $0$ 	& $1/4$	 & $-3/20$	& $\lambda^{-11/40}$\\
Middle w/o gap sat.	& $T_{c}^{\rm mos}$	&	$5.3(+5)$		& $-3/11$& $0$	& $-2/11$ & $-11/10$ 	 & $1/11$ & $2/11$	 & $61/110$	& $|\W|^{-26/55}$\\
Middle w/o gap uns.& $T_{c}^{\rm mou}$	&	$6.0(+5)$		& $-1$   	& $0$	& $0$       & $-11/10$ 	 & $1$    	& $1$	 & $1/10$		& $(\delta r_{\rm i}/r_{\rm H})^{-3}$\\
Near ext. sat.		& $T_{c}^{\rm nes}$	&	$1.6(+6)$		& $-3/11$& $0$	& $-2/11$& $-25/26$  	 & $1/11$ & $2/11$	 & $-6/11$	& $\delta\Omega^{1/26}\psi^{3/11}$ \\
Near ext. uns.		& $T_{c}^{\rm neu}$	&	$6.3(+5)$		& $-1$	& $0$	& $0$	& $-25/24$  	 & $1$	& $7/6$	& $1/24$		& $(\delta r/r_{\SCO})^{-1/2}\zeta_{\rm R}$ \\
Near int.			& $T_{c}^{\rm ni}$		&	$1.5(+6)$		& $0$	& $1/2$	& $-1/4$	& $-9/16$	 & $-1/4$	& $-1/2$	& $-5/16$	& $|\delta r/r|^{5/4}$ \\
\hline%
Far zone			&  $F_{0}$ 			& 	$7.9(13)$	& $0$	& $1$	& $-1$		& $-3$	  	& $0$ & $0$ & $0$ & $\varphi$\\
Middle with gap	& $F^{\rm mg}$		&	$1.8(15)$	& $1/2$	& $5/8$	& $-7/4$		& $-7/2$  	& $0$ & $5/8$	 & $-3/8$ & $\lambda^{-11/16}$\\
Middle w/o gap sat.	& $F^{\rm mos}$		&	$7.6(13)$	& $-2/11$& $0$	& $-21/22$    	& $-7/2$ 		& $5/22$	 & $5/11$	 & $61/44$ & $|\W|^{13/11}$\\
Middle w/o gap uns.& $F^{\rm mou}$		&	$1.0(14)$	& $-2$     & $0$    	& $-1/2$      	& $-7/2$  	& $5/2$ 	& $5/2$	 & $1/4$	 & $(\delta r_{\rm i}/r_{\rm H})^{-15/2}$\\
Near ext. sat.		& $F^{\rm nes}$		&	$9.8(14)$	& $-2/11$& $0$	& $-21/22$	& $-73/26$  	& $5/22$	 & $5/11$	& $-93/44$	 & $\delta\Omega^{5/26}\psi^{2/11}$ \\
Near ext. uns.		& $F^{\rm neu}$		&	$6.8(12)$	& $-2$	& $0$	& $-1/2$		& $-77/24$  	& $5/2$	 & $10/3$	& $-1/24$	 & $(\delta r/r_{\SCO})^{-5/2}\zeta_{\rm R}^{2}$ \\
Near int.			& $F^{\rm ni}$			&	$3.9(13)$	& $0$	& $1$	& $-1$		& $-7/4$ 		& $0$	 & $0$	& $-5/4$	& $|\delta r/r|$ \\
\hline%
Far zone			&  $|v_{r0}|$ 			& 	$3.6(+3)$		& $-4/5$	& $3/5$	& $1/5$	& $-2/5$	 & $0$ & $0$ & $0$ & $\varphi^{-3/5}$\\
Middle with gap	& $|v_r^{\rm mg}|$		&	$5.5(+2)$		& $1/2$	& $5/8$	& $1/4$	& $-1/10$ & $0$	 & $-3/8$	 & $9/40$ & $\lambda^{33/80}$\\
Middle w/o gap sat.	& $|v_r^{\rm mos}|$	&	$3.7(+3)$		& $9/22$& $1$	& $-5/22$& $-1/10$ & $-3/22$ &$-3/11$ & $183/220$	 & $|\W|^{-39/55}$\\
Middle w/o gap uns.& $|v_r^{\rm mou}|$	&	$3.1(+3)$		& $2$   & $1$	& $-1/2$ & $-1/10$ & $-3/2$ &$-3/2$ & $-3/20$	 & $(\delta r_{\rm i}/r_{\rm H})^{9/2}$\\
Near ext. sat.		& $|v_r^{\rm nes}|$		&	$6.6(+2)$		& $10/11$& $1$	& $-5/22$& $1/26$	 & $-3/22$& $-3/11$& $-41/44$	 & $\delta\Omega^{1/26}\psi^{-10/11}$ \\
Near ext. uns.		& $|v_r^{\rm neu}|$		&	$1.7(+2)$		& $2$	& $1$	& $-1/2$	& $-1/24$ & $-3/2$	& $-4/3$	& $-5/24$	 & $(\delta r/r_{\SCO})^{-1/2}\zeta_{\rm R}^{-2}$ \\
Near int.			& $|v_r^{\rm ni}|$		&	$3.0(+1)$		& $0$	& $0$	& $0$	& $-1/2$	 & $1$	& $2$	& $0$	& $|\delta r/r|^{-4}$ \\
\hline%
Far zone			&  $H_{0}^{\rm rad}$ 	& $1.5^{*}$ 	& $0$	& $1$	& $0$	& $0$		& $0$ & $0$ & $0$ & $\varphi$\\
Middle with gap	& $H_{\rm rad}^{\rm mg}$	& $35^{*}$		& $1/2$	& $5/8$	& $-3/4$	& $-1/2$ 	 & $0$	 & $1/8$	& $-3/8$	& $\lambda^{-11/16}$\\
Middle w/o gap sat.	& $H_{\rm rad}^{\rm mos}$	& $1.5^{*}$		& $-2/11$& $0$	& $1/22$& $-1/2$ 	 & $5/22$	 & $5/11$	& $61/44$& $|\W|^{13/11}$\\
Middle w/o gap uns.& $H_{\rm rad}^{\rm mou}$	& $1.9^{*}$		& $-2$   & $0$	& $1/2$	& $-1/2$ 	 & $5/2$	 & $5/2$	& $1/4$& $(\delta r_{\rm i}/r_{\rm H})^{-15/2}$\\
Near ext. sat.		& $H_{\rm rad}^{\rm nes}$	&	$19^{*}$	& $-2/11$& $0$	& $1/22$	& $5/26$ 	 & $5/22$	 & $5/11$	 & $39/44$& $\delta\Omega^{5/26}\psi^{2/11}$ \\
Near ext. uns.		& $H_{\rm rad}^{\rm neu}$	&	$0.13^{*}$	& $-2$ 	& $0$	& $1/2$	& $-5/24$ & $5/2$	 & $10/3$	 & $-1/24$ & $(\delta r/r_{\SCO})^{-5/2}\zeta_{\rm R}^{2}$ \\
Near int.			& $H_{\rm rad}^{\rm ni}$		&	$0.75^{*}$ 	& $0$	& $1$	& $0$	& $0$ 	& $0$	 & $0$	& $0$	& $|\delta r/r|$ \\
\hline%
Far zone			&  $H_{0}^{\rm gas}$ 	   & $0.28^{*}$    & $-1/10$& $1/5$	& $-1/10$	& $21/20$ & $0$ 	& $0$	& $0$ & $\varphi^{1/5}$\\
Middle with gap	& $H_{\rm gas}^{\rm mg}$	   & $0.53^{*}$	   & $0$	& $1/8$	& $-1/4$	& $19/20$ & $0$ 	& $1/8$	& $-3/40$& $\lambda^{-11/16}$\\
Middle w/o gap uns.& $H_{\rm gas}^{\rm mou}$  & $0.29^{*}$	   & $-1/2$	& $0$	& $0$	    & $19/20$ & $1/2$ 	& $1/2$	& $1/20$& $(\delta r_{\rm i}/r_{\rm H})^{-3/2}$\\
Near ext. uns.		& $H_{\rm gas}^{\rm neu}$  & $0.23^{*}$	   & $-1/2$ & $0$	& $0$		& $47/48$ & $1/2$	& $7/12$	& $1/48$ & $(\delta r/r_{\SCO})^{-1/4}\zeta_{\rm R}^{1/2}$ \\
Near int.			& $H_{\rm gas}^{\rm ni}$   & $0.36^{*}$   & $0$	& $1/4$ & $-1/8$	& $29/32$ & $-1/8$ & $-1/4$	& $-5/32$ & $|\delta r/r|^{5/8}$ \\
\hline
Far zone 			& $T_{\nu0}$			&	$7.1(47)$		& $0$	& $1$	& $2$		& $1/2$  	 & $0$	 & $0$	 & $0$	 & $\varphi$ \\
Middle with gap	& $T_{\nu}^{\rm mg}$	&	$1.6(49)$		& $1/2$	& $5/8$	& $5/4$		& $0$  	 & $0$	& $5/8$	& $-3/8$	 & $\lambda^{-11/6}$ \\
Middle w/o gap sat.	& $T_{\nu}^{\rm mos}$	&	$6.9(47)$		& $-2/11$& $0$	& $45/22$	& $0$  	 & $5/22$	& $5/11$	& $61/44$& $|\W|^{13/11}$ \\
Middle w/o gap uns.& $T_{\nu}^{\rm mou}$	&	$9.1(47)$		& $-2$	& $0$	& $5/2$		& $0$  	 & $5/2$	 & $5/2$	 & $1/4$	& $b^3$ \\
Near ext. sat.		& $T_{\nu}^{\rm nes}$	&	$1.1(49)$		& $-2/11$& $0$	& $45/22$	& $0$  	 & $5/22$	 & $5/11$	& $61/44$& $\psi^{13/11}$ \\
Near ext. uns.		& $T_{\nu}^{\rm neu}$	&	$1.7(49)$		& $-2$	& $0$	& $5/2$		& $0$  	 & $5/2$	 & $5/2$	 & $1/4$	& $\zeta_{\rm R}^3$ \\
Near int.			& $T_{\nu}^{\rm ni}$	&	$2.4(50)$		& $1$	& $3/2$	& $7/4$		& $15/16$ 	 & $-5/4$	& $-5/2$	 & $-5/16$	& $|\delta r/r|^{21/4}$ \\
\hline
GW inspiral 	& $|v_{\SCO r,\rm GW}|$ & $380$ & $0$ & $0$ & $0$  & 0 & $0$  & $1$   & $-3$   & \\
Type-II 		& $|v_{\SCO r,II}|$       & $550$ & $1/2$    & $5/8$   & $1/4$   & 0 & $0$ & $-3/8$ & $1/8$ & $\lambda^{-11/16}$\\
Type-1.5 sat.	& $|v_{\SCO r,1.5\rm s}|$ & $23$ & $-2/11$    & $0$   & $23/22$   & 0 & $5/22$ & $-6/11$ & $83/44$ & $|\W|^{13/11}$\\
Type-1.5 uns.	& $|v_{\SCO r,1.5\rm u}|$ & $31$ & $-2$       & $0$   & $3/2$     & 0 & $5/2$  & $3/2$   & $3/4$   & $b^3$\\
\hline
Middle gap uns.	& $k^{\rm mgu}_{\SCO}$  & $23$   & $1/2$    & $-3/8$    & $-3/4$    & $0$   & $0$       & $5/8$     & $-7/8$    & $\lambda_{\rm u}^{-11/16}$\\
Middle gap sat. 	& $k^{\rm mgs}_{\SCO}$  & $23$   & $1/2$    & $-3/8$    & $-3/4$    & $0$   & $0$       & $5/8$     & $-7/8$    & $\lambda_{\rm s}^{-11/16}$\\
Middle w/o gap uns	& $k^{\rm mou}_{\SCO}$ & $1.3$  & $-2$     & $-1$      & $1/2$     & $0$   & $5/2$     & $5/2$     & $-1/4$    & $b^3$\\
Middle w/o gap sat.	& $k^{\rm mos}_{\SCO}$ & $0.97$ & $-2/11$  & $-1$      & $1/22$    & $0$   & $5/22$    & $5/11$    & $39/44$   & $|\W|^{13/11}$\\
Near ext. sat. 		& $k^{\rm nes}_{\SCO}$   & $5.5$  & $-4/17$  & $-1$      & $1/17$    & $0$   & $5/17$    & $10/17$   & $29/34$   & $|\W|^{4/17}$\\
\hline
argument of $\W$ & $-a$			& $-0.465$ 		& $1$	& $0$	& $-1/4$		& $0$	& $-5/4$	& $-13/12$ & $5/8$ & $ (\delta r_{\rm i}/r_{\rm H})^{17/4}$\\
\hline\hline
\end{tabular}
\caption{\label{t:analytical} Pre-factors and exponents in the analytical disk model in different zones, $C$ and $c_i$ in Eq.~(\ref{e:X}).  The third column is in cgs units except where marked by * (where it is in units of $\G M_{\SMBH}/\C^2$). Columns 4--10 are exponents, the last column is the extra multiplicative function (see text). The last two block of parameters show the migration rate of the secondary and other useful parameters.
}
\end{table*}

\begin{table*}
\begin{tabular}{lcrccccccc}
\hline\hline
			&						&	$C$		& $[\alpha_{-1}]$& $[\dot{m}_{-1}]$& $[M_7]$	& $[f_{-2}]$ & $[q_{-3}]$	& $[r_{\SCO 2}]$  & $\Phi$\\
\hline
Middle/Far w. gap sat.		& $r_{\rm f}^{\rm mgs}/r_{\SCO}$		& $540$		& $1$		& $-3/4$	& $-3/2$		& $0$		& $5/4$  	   & $-7/4$  &  $(1+\delta_{\rm mg}^{\rm nes})^{-11/8}$\\
Middle/Far w. gap uns.		& $r_{\rm f}^{\rm mgu}/r_{\SCO}$		& $540$		& $1$		& $-3/4$	& $-3/2$		& $0$		& $5/4$  	   & $-7/4$  &  $(1+\delta_{\rm mg}^{\rm neu})^{-11/8}$\\
Near ext./middle w. gap sat.    & $\delta_{\rm mg}^{\rm nes}$	& $0.130$ 	& $-4/37$	& $4/37$	& $-3/37$	& $9/37$	& $22/37$  & $-33/74$  &  \\
Near ext./middle w. gap uns.    & $\delta_{\rm mg}^{\rm neu}$	& $0.55$		& $-4/37$	& $-6/37$& $-3/37$	& $9/37$	& $22/37$  & $-13/74$   &  $(1+\delta_{\rm mg}^{\rm neu})^{-35/148}$\\
Near ext./middle w/o gap sat.   & $\delta_{\rm mo}^{\rm nes}$	& $0.083$		& $-4/17$	& $0$& $1/17$	& $5/17$	& $10/17$  & $-5/34$  & $|\W|^{4/17}$  \\
Near ext./middle w/o gap uns.   & $\delta_{\rm mo}^{\rm neu}$	& $0.35$		& $-4/7$	& $-2/7$& $1/7$	& $5/14$	& $20/21$  & $-1/14$  & $(r_{\rm i}/r_{\SCO})^{-115/126}$  \\
Far/Near int. 		& $\delta_{\rm ni}^{\rm f}$		& $-0.1$		& $-4/17$		& $-2/17$	& $1/17$		& $5/17$	& $10/17$	& $-1/34$  &  $(1+\delta_{\rm ni}^{\rm f})^{-841/714}$\\
\hline\hline
\end{tabular}
\caption{\label{t:analytical-radii}
 Transition radii between different zones relative to the secondary orbital radius, $r_{\SCO}$.
 Here $r_{a}^{b}$ is the radius at the interface between zone $a$ and $b$,
$\delta r_{a}^{b} \equiv r_{a}^{b}-r_{\SCO}$, and $\delta_{a}^{b} = \delta r_{a}^{b} /r_{\SCO}$.
Different columns show the constant prefactor and exponents in Eq.~(\ref{e:X}) as in Table~\ref{t:analytical}.
}
\end{table*}

\subsection{Disk model}

Our results in this paper apply to geometrically thin, optically thick
accretion disks, and describe vertically and azimuthally averaged
properties.  All physical parameters can be written as
\begin{equation}\label{e:X}
 X(r,r_{\SCO},\mathbf{p}) = C\, \alpha_{-1}^{c_1}\, \dot{m}_{-1}^{c_2}\, M_7^{c_3}\, r_{2}^{c_4}\, f_{-2}^{c_5} \, q_{-3}^{c_6} \, r_{\SCO 2}^{c_7} \,\Phi(r,r_{\SCO},\mathbf{p})
\end{equation}
where $X$ denotes any of $\{\Sigma, T_{c}, H, v_r, F, T_{\nu}\}$;
$r_{2}$ and $r_{\SCO 2}$ denote the radius from the primary and the
orbital radius of the secondary in units $100\, M_{\SMBH}$,
respectively; and $\Phi(r,r_{\SCO},\mathbf{p})$ denotes an extra
function of the parameters $\mathbf{p}=(\alpha,\dot{m}, f, M, q)$.
Note that $(\alpha,\dot{m}, M)$ are the usual parameters of a standard
solitary accretion disk; $q$ and $f$ represent the mass ratio and the
normalization of the azimuthally averaged tidal torque (see Eq.~\ref{e:Lambda}).
The $_{-N}$ index denotes normalizing with $10^{-N}$,
e.g. $q_{-3}=q/10^{-3}$.  The $C$ prefactor, $c_i$ exponents, and the
$\Phi$ function in Eq.~(\ref{e:X}) are given in
Table~\ref{t:analytical} in the different zones and cases for
$\beta$-disks (i.e. $\nu\propto p_{\rm gas}$).  For $\beta$-disks,
many of the physical parameters, including the surface density,
$\Sigma$, and the central temperature, $T_{\rm c}$, are independent of
whether gas or radiation pressure dominates. This, however, is not
true for the scale-height, where we quote results in both regimes,
labeled with a ``gas'' or ``rad'' subscript.

The last column of Table~\ref{t:analytical} shows
$\Phi(r,r_{\SCO},\mathbf{p})$ in Eq.~(\ref{e:X}).  Here we introduced
the following notation:
\begin{align}
\delta r&=r-r_{\SCO}\,,\quad \delta r_{\rm i}=r_{\rm i}-r_{\SCO}\,,\\
\lambda_{\rm s} &= 1 + \delta_{\rm mg}^{\rm nes}\,,\quad\lambda_{\rm u} = 1 + \delta_{\rm mg}^{\rm neu}\,,\label{e:results-lambda}\\
 \delta \Omega &\equiv (\Omega_{\SCO}-\Omega)/\Omega_{\SCO}=1 - (r/r_{\SCO})^{-3/2}\,,\\
 \delta \Omega_0 &\equiv (\Omega_{\SCO}-\Omega_0)/\Omega_{\SCO}=1 - (r_{\rm i}/r_{\SCO})^{-3/2}\,,\\
 \varphi &\equiv 1 - \left(\frac{r_{\ISCO}}{r}\right)^{1/2}\,,\\
 \zeta_{\rm R} &\equiv
\frac{r_{\rm H}^{5/2}}{r_{\SCO}^{5/2}}
\int_{r_{\rm i}/r_{\SCO}}^{r/r_{\SCO}} x^{-7/6} \left(1-x^{-3/2}\right)^{-1/6}(x - 1)^{-10/3}\,\D x\nonumber\\
&\approx \frac{2^{1/6}}{3^{1/6}}\frac{2}{5}
\frac{(r_{\rm i}/r_{\SCO})^{-115/72}}{(\delta r_{\rm i}/r_{\rm H})^{5/2}}
\left[1 - \left(\frac{r_{\rm i}}{r}\right)^{115/72}\left(\frac{\Delta_{\rm i}}{\Delta}\right)^{5/2}\right]\,,\\
\psi &\equiv \frac{2}{3}\int_{\delta \Omega}^{\delta \Omega_0} x^{-21/26} (1-x)^{-6/13} \D x\nonumber\\
&=\frac{2}{3}\Beta\hspace{-2pt}\left(\delta\Omega; 5/26 , 7/13 \right)
-\frac{2}{3}\Beta\hspace{-2pt}\left(\delta\Omega_{0}; 5/26, 7/13\right)\nonumber\\
&\approx 0.4 \ln \frac{\delta r}{\delta r_{\rm i}}\,,\\
\W &\equiv \W_{-1}(-a) \approx \ln(a)-\ln(-\ln(a))\,,\\
 b&\approx \left(\frac{r_{\rm i}}{r_{\SCO}}\right)^{-115/72} \left(\frac{\delta r_{\rm i}}{r_{\rm H}}\right)^{-5/2}.
\end{align}
Here $B(x; a, b)$ is the incomplete Beta function, and $\W$ is the
Lambert-W function, the branch defined on the negative real axis,
evaluated at $-a$ given in the last row of Table~\ref{t:analytical}.
Typically, $1 \leq |\W(-a)| \lesssim 10$.  The approximations shown
for $\psi$ and $\W$ are better than $20\%$.  Here, $r_{\ISCO}$ is the
innermost stable circular orbit near the SMBH (i.e. $6\, M_{\SMBH}$
($1\,M_{\SMBH}$) for a non-spinning (maximally spinning) SMBH)),
$r_{\rm H}=(q/3)^{1/3}r_{\SCO}$ is the Hill radius, $r_{\rm i}$ marks
the radius at which the viscous torque becomes very small outside the
secondary's orbit for which we use
\begin{equation}
\delta r_{\rm i} = r_{\rm H} {~~~\rm if~~} k_{\SCO}=k_{\SCO}^{\rm mou} {\rm ~~or~~} k_{\SCO}=k_{\SCO}^{\rm mos},
\end{equation}
and $\lambda$ is the dimensionless gap { or truncation} radius scale
\begin{equation}
\lambda =
\left\{
\begin{array}{ll}
\lambda_{\rm u}       &{\rm if~} k_{\SCO}=k_{\SCO}^{\rm mgu}\,, \\
\lambda_{\rm s}       &{\rm if~} k_{\SCO}=k_{\SCO}^{\rm mgs}\,,
\end{array}
\right.
\end{equation}
where\footnote{If either $k_{\SCO}^{\rm mgu}$, $k_{\SCO}^{\rm mgs}$,
$k_{\SCO}^{\rm mou}$, or $k_{\SCO}^{\rm mos}$ is less than one, then the disk does
not have a middle zone by definition, and the disk is not strongly
perturbed.}
\begin{equation}
k_{\SCO}=\{k_{\SCO}^{\rm mou},k_{\SCO}^{\rm mos},k_{\SCO}^{\rm mgu},k_{\SCO}^{\rm mgs}\}
\end{equation}
is the { dimensionless angular momentum flux or } {\emph{brightening factor}}.

\begin{figure*}
\centering
\mbox{\includegraphics[width=8.5cm]{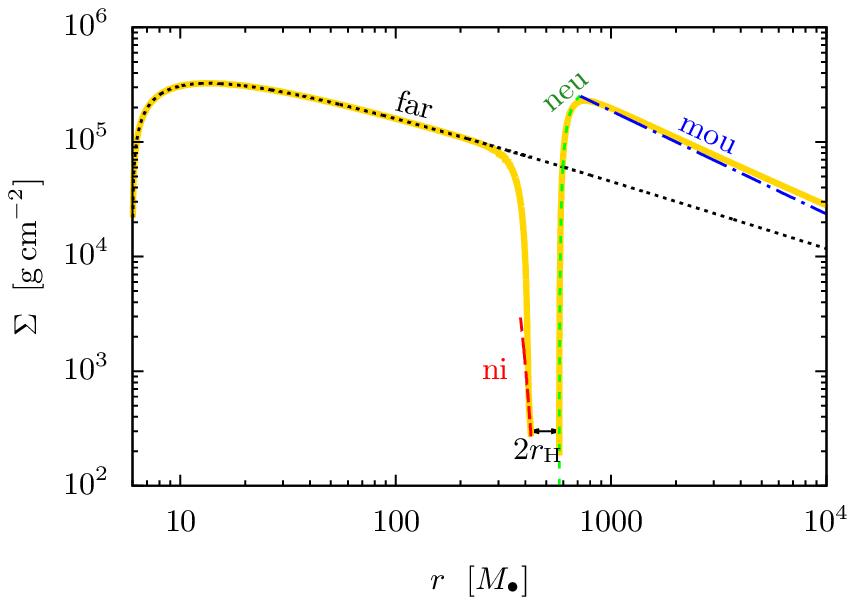}}
\mbox{\includegraphics[width=8.5cm]{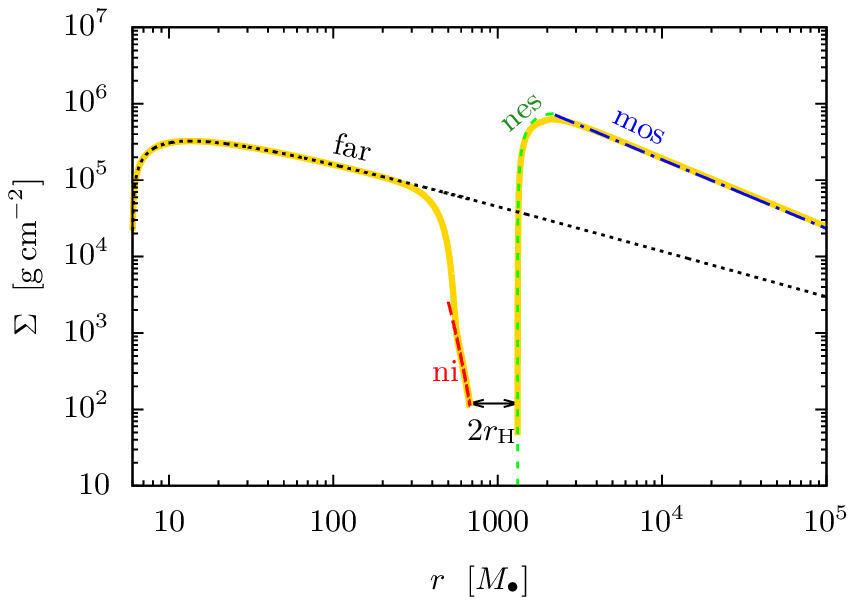}}
\caption{\label{f:migration-comp}
Comparison of the numerical solution (thick yellow solid lines) with
the asymptotic analytical approximations in the various zones (dotted,
dashed, and dash-dotted lines, labeled as in Table~\ref{t:analytical})
for the surface density of the disk around a $10^5\,\Msun$
primary. The mass ratio and binary separation are $(q,r_{\SCO}) =
(0.01, 500\,M)$ and $(0.1, 1000\,M)$ on the left and right panels,
respectively. In both cases the disk is in the overflowing
steady-state.  The tidal torque is unsaturated in the near zone
outside the Hill radius on the left panel, but it is saturated on the
right panel. The disk structure is significantly modified in both
cases in an extended region around the secondary.  }
\end{figure*}

We collect the formulae for the transition radii separating different
radial zones and physical regimes in
Table~\ref{t:analytical-radii}. We label
$\delta^{b}_{a}=(r^{b}_{a}-r_{\SCO})/r_{\SCO}$, where $r^{b}_{a}$
marks the radius of the interface between zone $a$ and $b$.  For
example, $\delta_{\rm mg}^{\rm nes}$ is the transition between the
middle zone and the torque-saturated exterior near zone if there is a
{ wide} gap, which also sets the { truncation radius} scale $\lambda_{\rm s}$ in
Eq.~(\ref{e:results-lambda}).

The state of the disk and the migrate speed of the binary is directly
set by the { dimensionless angular momentum flux} $k_{\SCO}$. If
$k_{\SCO}=$[$k_{\SCO}^{\rm mou}$]$k_{\SCO}^{\rm mos}$, then the disk
is in the [un]saturated overflowing state, whereas if
$k_{\SCO}=$[$k_{\SCO}^{\rm mgu}$]$k_{\SCO}^{\rm mgs}$ then it is in
the [un]saturated state with a { wide} gap.  These four possibilities are
indicated in Table~\ref{t:analytical}.  The appropriate choice of
$k_{\SCO}$ also determines which case in Table~\ref{t:analytical} are
to be used in the middle zone for the other parameters ('mou', 'mos',
or 'mg'), and in the near exterior zone ('neu' or 'nes').

Thus, the ``phase space'' of solutions consists of four
regions\footnote{Here we assume that GW emission is negligible, and
the disk drives the binary. See Paper~II for further discussion.}.
The transition between two different solutions, with or without { wide} gaps,
corresponds to the parameters for which $k_{\SCO}^{\rm
mos}=k_{\SCO}^{\rm mgs}$ or $k_{\SCO}^{\rm mou}=k_{\SCO}^{\rm mgu}$.
Our hypothesis is that this must represent a physical transition in
the disk+binary system, as the secondary migrates inward from large
radius. Initially, a central cavity is created, and the outer edge of
the cavity lies far from the secondary's orbit.  However, as the
secondary migrates inward, the distance between the cavity edge and
the secondary shrinks (at least when measured in units of the Hill
radius of the secondary).  This may happen both because the viscosity
increases as the pressure grows during pile-up, and also because the
tidal torque decreases with increasing scaleheight due to the torque
cutoff.  The cavity finally closes once the cavity wall nudges inside
the Hill radius.  The { dimensionless angular momentum flux or} brightening factor, $k_{\SCO}$, is largest when
this transition occurs.

These gap opening/closing conditions are quite different from those in
the literature \citep[e.g.][]{1986Icar...67..164W}, which state that
the gap closes if the disk extends into the region closer than either
the local scaleheight or the Hill radius from the secondary.  Note
that our gap closing condition combines statements on the scaleheight
and the Hill radius, but since the scaleheight and viscosity vary
significantly near the secondary in the strongly perturbed case with a
large pile-up, they depend on the actual perturbed profiles rather
than the averaged quantities describing accretion disks around a
solitary object.  We discuss gap opening/closing, and its physical
implications, in more detail in Paper~II.

We have verified that the analytical approximate solutions to the disk
model match the numerical solutions typically to within tens of
percent for a wide range of disk and binary parameters when the disk
is strongly perturbed.  Figure~\ref{f:migration-comp} shows two
examples when the disk is in the overflowing state with unsaturated
(left panel) and saturated (right panel) tidal torques near the
secondary. Different regions are indicated with the abbreviations used
in Table~\ref{t:analytical}.  For further examples and other physical
quantities, see Paper~II.

\subsection{Brightening factor}

As stated in the previous sections, the $k_{\SCO}$ parameter sets the
{ angular momentum flux and the }
brightness of the disk in the middle zone relative to the unperturbed
value, and determines the state of the disk.

Remarkably, the disk parameters can differ dramatically from the
unperturbed values not only in the near zone, where the tidal effects
dominate, but also in the middle zone, where tidal effects are already
negligible.  The tidal effects of the secondary are short-range, but
the corresponding effect is communicated to distant regions by setting
an effective boundary condition. The size of this region can be
$r^{\rm mg}_{\rm f}/r_{\SCO}\sim 540\, \alpha_{-1}\dot{m}_{-1}^{-3/4}
M_7^{-3/2}r_{\SCO 2}^{-7/4} q_{-3}^{5/4}$ times larger than the
secondary orbital radius (see Table~\ref{t:analytical-radii}).  This
long-range behavior is confirmed in our full numerical solutions and
is also present in \citet{2009MNRAS.398.1392L} and
\citet{2010MNRAS.407.2007C} for a circumbinary disk with a { cavity}.  The
same $k_{\SCO}$ parameter also sets the increase in the scale-height,
as well as the migration speed of the secondary.  We have derived the
analytical formulae for $k_{\SCO}$ in the four relevant regimes
summarized in the previous section.  Due to its extended radial size,
the integrated luminosity of the middle zone is often larger than that
of the near zone in the torque-unsaturated case, overflowing state.
However, the near zone may be brighter than the middle zone, when the
system is in the torque-saturated, overflowing state (see Paper~II for
further discussions).

\subsection{Migration rate}

\begin{figure}
\centering
\mbox{\includegraphics[width=8.5cm]{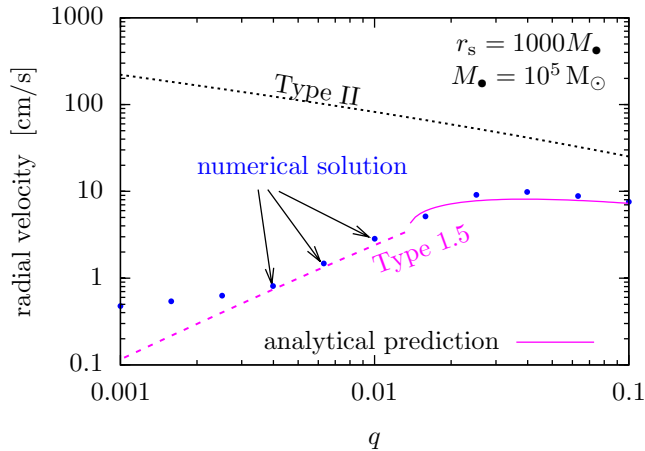}}
\caption{\label{f:migrationrate-big}
The migration speed of the secondary for different mass ratios for
$M_{\SMBH}=10^5\Msun$ at $r_{\SCO}=1000 M_{\SMBH}$ (or $t_{\rm
orb}=1\,\rm day$).  The steady-state numerical solutions (blue) are
well represented by our analytical formula for Type-1.5 migration
(magenta), but not by those for Type-I or II (black dotted).
The dashed and solid magenta lines show the unsaturated and
saturated Type-1.5 cases, respectively. The discrepancy between
the numerical and analytical solutions become more significant at $q<0.002$
due to the fact that the disk is not strongly perturbed there.
 }
\end{figure}

Finally, we have derived the migration speed of the secondary;
approximate analytical formulae for our results are listed in
Table~\ref{t:analytical}.  We find that in the regimes when a { cavity
forms}, the migration in our solution is slower than the standard
steady-state secondary-dominated Type-II migration rate
\citep{1995MNRAS.277..758S} by a factor $\lambda^{11/16}\sim 2$.  The
Type-II rate may be further reduced in non-steady state models if the
accretion rate or the gas mass is limited
\citep{1999MNRAS.307...79I,2009MNRAS.398.1392L}.  However, in the new
``Type-1.5'' regime we identify, with a partial pile-up and overflow,
the migration is even slower.

More generally, the migration speed for a strongly perturbed disk
follows the minimum of the three possible solutions:
\begin{equation}
|v_{\SCO r}| = \min( |v_{\SCO r, 1.5\rm s}|, |v_{\SCO r, 1.5\rm u}|, |v_{\SCO r, II}|)\,.
\end{equation}

In general, Type-1.5 migration is more rapid at larger $r_{\SCO}$, in
contrast with the Type-II rate, which is nearly constant, and the
GW inspiral rate, which strongly decreases with $r_{\SCO}$. Therefore,
as a real system evolves, it will first transition from an initial
Type-II migration at large radii to Type-1.5 migration at smaller
radii, before finally being driven by GWs at still smaller
separations.

Regarding the dependence on the primary mass, the Type-II is nearly
constant, while the Type-1.5 speed increases with $M_{\SMBH}$.  This
implies that at any fixed orbital separation, Type-1.5 migration is
relevant for lower masses (roughly those in the range expected to be
detectable by a space-based gravitational wave mission such as {\it
LISA}, and Type-II is relevant for higher-mass binaries (in the
sensitivity range of Pulsar Timing Arrays; see Paper II).  Type-1.5
migration may also be important for stellar mass binaries in
proto-stellar disks or Jupiter-mass planets around M-type dwarf stars
of mass (0.1--$1)\,\Msun$ \citep[see][for a recent discovery of such a
system]{2012AJ....143..111J}.  Finally, we note that Type-1.5
(Type-II) migration operates for larger (smaller) accretion rates, if
fixing all other parameters.

All of the above conclusions are based on the analytical solutions we
obtained; however, we have verified, by numerically solving the
equations presented in \S~\ref{s:interaction}, that our solutions are
accurate to within tens of percent for a broad range of parameters.
In particular, in Figure~\ref{f:migrationrate-big} we show the the
analytical approximation of the migration rates for the case of
$r_{\SCO}=1000\,M_{\SMBH}$ for $M_{\SMBH}=10^5\Msun$, together with
the rates obtained from a numerical solution.  In this case, a disk is
in the overflowing state for all values of $q\gtrsim 10^{-3}$ shown,
and the migration rate is significantly different from both Type-I and
II. As the figure shows, the analytical Type-1.5 formulas give a good
approximation over a wide range of $q$ for the migration speed.

\subsection{Caveats}

Our findings are subject to many possible caveats.
\begin{itemize}
 \item We assumed a radiatively efficient disk model in which the
effective viscosity is proportional to the gas pressure in the disk
with a constant $\alpha$ coefficient even in the radiation pressure dominated regime. Future studies should
investigate alternative models in which the viscosity is proportional
to the total gas+radiation pressure \citep{1973A&A....24..337S}, or
where the viscosity is generated by magneto-rotational instability
(MRI, see
\citealt{2003ApJ...593..992T,2012ApJ...749..118S,2012arXiv1203.6108G,2012arXiv1204.1073N}
for simulations of circumbinary disks leading to an ``antigap'').
\item We assumed steady-state models where the accretion rate is
constant over radius.  This is expected to be valid as long as the gas
inflow is much faster than the migration rate of the secondary { over a large range of radii}, if the
total gas supply is not limited, and if the accretion rate is set at
the inner or outer boundary (i.e. for the new Type-1.5 migration regime
we focus on here).  However, this assumption may be violated for
{ tidally truncated circumbinary disks} \citep[Type-II
migration,][]{1999MNRAS.307...79I}
{ or for models where the gas supply rate is limited at the outer boundary
\citep{2009MNRAS.398.1392L,2012arXiv1205.5017R}}.
 \item We assumed unequal-mass binaries, averaged over the azimuthal
angle, assumed that the density waves generated by secondary are
dissipated locally, and that the radial tidal torque profile follows
the formula given by \citet{2002ApJ...567L...9A}.  This assumes that
the tidal torque saturates near the secondary at a radial distance
closer than the scaleheight ($\partial_r T_{\rm d}\propto H^{-4}$).
We also assumed that the gas entering a distance comparable to the
Hill radius can flow freely across the secondary's orbit.
{ However, accretion onto the secondary may affect the Type-1.5 migration rate.}
Farther
away from the secondary, the assumed torque density has a steep cutoff
($\partial_r T_{\rm d}\propto |r-r_{\SCO}|^{-4}$); extrapolating
beyond $r>2r_{\SCO}$ might be inaccurate.
{ The $|r-r_{\SCO}|^{-4}$ scaling may also be inaccurate in the local nonlinearly
perturbed regime especially for comparable mass-ratio binaries
\citep{2008ApJ...672...83M,2012arXiv1202.6063R,2012arXiv1203.5798P}.}
 These issues should be
investigated using simulations, which could also address comparable
mass binaries where the disk may be significantly non-axisymmetric
\citep[e.g.][]{2008ApJ...672...83M,2009MNRAS.393.1423C} { and where the
accretion of the secondary is non-negligible
\citep{1999ApJ...526.1001L}.}
 \item We neglected non-axisymmetric inflow into the secondary's orbit { or onto the secondary}
if a { cavity} is formed.  Inflow across the gap { or accretion of the secondary} reduces the amount of pile
up outside the secondary, reduces the Type-II migration rate, and
could affect the gap closing transition between the continuously
overflowing solutions and the cases with a gap.  { We have also neglected the corotation torques in the
overflowing case.}
 \item We found that the enhanced pressure dominates over the increase
in surface density outside the secondary's orbit, which makes the overflowing disk stable
against gravitational fragmentation (see Paper~II).  However, the steep pressure
gradient in the near zone around a massive secondary may lead to
global non-axisymmetric dynamical instabilities \citep{1985MNRAS.213..799P,1986MNRAS.221..339G}.
The corresponding enhancement of the effective viscosity and angular momentum transport in the disk
might reduce the pile-up outside the secondary's orbit and
further reduce the Type-1.5 migration rate. A detailed stability analysis and an investigation of its implications
goes beyond the scope of this paper.
 \item We restricted our attention to circular binaries. However,
binary eccentricity may be excited in cases with a { cavity}, when the
masses of the two compact objects are comparable and the gap edge
itself becomes significantly non-axisymmetric
\citep{1992PASP..104..769A,2005ApJ...634..921A,2011MNRAS.415.3033R,2012arXiv1202.6063R}.
We note that such non-axisymmetries excited in disks with a gap
diminish once the mass ratio is $q\lesssim 0.1$ (D'Orazio et al. 2012, in
preparation).  Nevertheless, it remains to be seen if the binary
develops significant eccentricities in the unequal-mass, overflowing
state, with a significant pile-up.
\item We assumed that the binary is sufficiently widely separated that
gravitational wave emission is negligible.  For a complete picture,
future studies should investigate the gravitational wave driven regime
\citep{2010ApJ...714..404T,2010MNRAS.407.2007C,2011PhRvL.107q1103Y,2011PhRvD..84b4032K,2011PhRvD..84b4024F,2012MNRAS.420..705T,2012ApJ...744...45B,2012MNRAS.tmpL.436B,2012arXiv1203.6108G,2012arXiv1204.1073N}.
\end{itemize}

\section{Conclusions}\label{s:conclusions}

In summary, we have presented new analytical solutions to disk
properties and migration rates, obtained from self-consistent
solutions of a coupled binary-disk system. The evolution equations are
solved analytically in the strongly perturbed limit, including the
angular momentum exchange between the disk and the binary and the
modifications to the density, scale-height, and viscosity
self-consistently, including viscous and tidal heating, diffusion
limited cooling, radiation pressure, and the orbital decay of the
binary.

In addition to recovering solutions with a central cavity, similar to
previous ``Type-II migration'' scenarios, we have identified a
distinct new regime, applicable at smaller separations and masses,
larger accretion rates, and mass ratios in the range $ 10^{-3}\lesssim
q \lesssim 0.1$.  For these systems, gas piles up outside the binary's
orbit, but rather than creating a cavity, it continuously overflows as
in a porous dam.  { The disk properties are intermediate between 
those in an unperturbed disk and a disk with a wide gap.}
The migration rate of the secondary in this ``Type
1.5'' regime is typically slower than both Type-I and Type-II rates.

In this paper, we have presented simple analytical formulae that
comprehensively describe binary systems with different parameters, in
various stages of evolution.  The analytical results provide simple
scaling relations, which may be useful to scale and interpret the
results of numerical simulations to different disk or binary
parameters.  It allows us to map out the effects of varying $\alpha$,
$\dot{m}$, or the binary parameters, over a wide range from planetary
disks to active galactic nuclei around SMBH binaries.

We discuss the applicability of the new Type-1.5 regime and its
physical implications for specific systems, as well as possible
observable signatures in a companion paper \citep{paper2}.

\section*{Acknowledgments}

We thank Re'em Sari, { Taka Tanaka, Alberto Sesana, and Roman Rafikov} for useful discussions.
BK acknowledges support from NASA through Einstein Postdoctoral
Fellowship Award Number PF9-00063 issued by the Chandra X-ray
Observatory Center, which is operated by the Smithsonian Astrophysical
Observatory for and on behalf of the National Aeronautics Space
Administration under contract NAS8-03060.  This work was supported in
part by NSF grant AST-0907890 and NASA grants NNX08AL43G and
NNA09DB30A (to AL) and NASA grant NNX11AE05G (to ZH).

\appendix
\section{Viscously and tidally heated disks}\label{s:app:thermal}
Here we provide the details of the algebraic manipulations that give
the disk model for fixed $F$ and $D_{\nu}$ for either $\alpha$ or
$\beta$--disks (i.e. $b=0$ or 1).

For fixed $\beta$, we can reduce the problem to 3 equations and 3
unknowns $\Sigma, T, \nu$. From Eqs.~(\ref{e:Fdef}), (\ref{e:nu}), and
(\ref{e:Dnu}),
\begin{align}
 \Sigma \nu &= \frac{8}{9\Omega^2}D_{\nu} =  \alpha  \frac{\kappa^2}{\Omega^3} \frac{ \beta^{b} F^2}{(1-\beta)^2} \Sigma \\
F &=  \frac{8}{3} \frac{\sigma}{\kappa} \frac{T_c^4}{\Sigma}
\end{align}
Solve this for $\Sigma$ and $T$,
\begin{align}\label{e:Sigma1}
 \Sigma &= \frac{8 \C^2}{9\alpha\kappa^2}\Omega\frac{(1-\beta)^2}{\beta^{b}}\frac{D_{\nu}}{F^2}\\
T_c^4 &= \frac{\C^2 }{3 \alpha\kappa\sigma}\Omega\frac{(1-\beta)^2}{\beta^{b} }\frac{D_{\nu}}{F}
\label{e:Tc1}
\end{align}
Finally $\beta$ is given by
\begin{align}
 \frac{\beta}{1-\beta} &= \frac{p_{\rm gas}}{p_{\rm rad}} = \frac{3 \rho k T_c}{a \mu m_p T_c^4}=\frac{3 k}{ a \mu m_p}\frac{\Sigma}{2 H T_c^3}
\end{align}
where $\Sigma$, $T_c$, and $H$ are to be substituted from
Eqs.~(\ref{e:Sigma1}), (\ref{e:Tc1}), and (\ref{e:H}).  From this
\begin{align}\label{e:beta1}
 &\frac{\beta^{(1/2) + (b-1)/10}}{1-\beta} =
\frac{\C [k /(\mu m_p)]^{2/5} }{(3\,\alpha \sigma)^{1/10} \kappa^{9/10}} \Omega^{9/10} \frac{D_{\nu}^{1/10}}{F^{9/10}}
\end{align}
Finally Eqs.~(\ref{e:Sigma1}) and (\ref{e:Tc1}) can be simplified by
eliminating $1-\beta$ using Eq.~(\ref{e:beta1}), which leads to
Eq.~(\ref{e:Sigma}--\ref{e:Tc}).

\bibliography{paper-1}

\end{document}